\documentclass[aps,prd,nofootinbib,amssymb,eqsecnum,notitlepage]{revtex4-1}
\usepackage{graphicx}
\usepackage{bm}
\usepackage{amsmath,amsthm,amssymb}
\usepackage{color}

\usepackage{amsfonts}
\usepackage{hyperref}
\usepackage{comment}

\begin{document}
\newcommand{\newc}{\newcommand}

\newc{\be}{\begin{equation}}
\newc{\ee}{\end{equation}}
\newc{\ba}{\begin{eqnarray}}
\newc{\ea}{\end{eqnarray}}
\newc{\bea}{\begin{eqnarray*}}
\newc{\eea}{\end{eqnarray*}}
\newc{\Mpl}{M_{\rm pl}}
\newc{\da}{\delta{A}}
\newc{\mm}[1]{\textcolor{red}{#1}}
\newcommand{\mmc}[1]{\textcolor{blue}{[MM:~#1]}}

\allowdisplaybreaks[1]

\title{Dark energy scenario consistent with GW170817 
in theories beyond Horndeski gravity}

\author{Ryotaro Kase}

\affiliation{Department of Physics, Faculty of Science,
Tokyo University of Science, 1-3,
Kagurazaka, Shinjuku, Tokyo 162-8601, Japan}

\author{Shinji Tsujikawa}

\affiliation{Department of Physics, Faculty of Science, 
Tokyo University of Science, 1-3,
Kagurazaka, Shinjuku, Tokyo 162-8601, Japan}

\begin{abstract}

The Gleyzes-Langlois-Piazza-Vernizzi (GLPV) theories up to quartic order are the general scheme of scalar-tensor theories allowing the possibility for realizing the tensor propagation speed $c_t$ equivalent to 1 on the isotropic cosmological background. We propose a dark energy model in which the late-time cosmic acceleration occurs by a simple k-essence Lagrangian analogous to the ghost condensate with cubic and quartic Galileons in the framework of GLPV theories. 
We show that a wide variety of the variation 
of the dark energy equation of state $w_{\rm DE}$ including the 
entry to the region $w_{\rm DE}<-1$ can be realized without violating 
conditions for the absence of ghosts and Laplacian instabilities. 
The approach to the tracker equation of state $w_{\rm DE}=-2$ 
during the matter era, which is disfavored by observational data, can be 
avoided by the existence of a quadratic k-essence Lagrangian $X^2$. 
We study the evolution of nonrelativistic matter perturbations for 
the model $c_t^2=1$ and show that the two quantities $\mu$ and 
$\Sigma$, which are related to the Newtonian and weak lensing 
gravitational potentials respectively, are practically equivalent to 
each other, such that $\mu \simeq \Sigma>1$.
For the case in which the deviation of $w_{\rm DE}$ 
from $-1$ is significant at a later cosmological epoch, 
the values of $\mu$ and $\Sigma$ tend to be larger 
at low redshifts.
We also find that our dark energy model can be consistent with 
the bounds on the deviation parameter $\alpha_{\rm H}$ from 
Horndeski theories arising from the modification of gravitational law inside massive objects.

\end{abstract}
\date{\today}


\maketitle

\section{Introduction}
\label{intro} 

Two decades have passed after the first observational discovery 
of cosmic acceleration by the supernovae 
type Ia (SN Ia) \cite{SN1,SN2}.
With the SNIa data alone, the equation of state of dark energy 
$w_{\rm DE}$ has not been strongly constrained. 
However, the joint data analysis combined with the observations 
of Cosmic Microwave Background 
(CMB) \cite{WMAP,Planck} and Baryon Acoustic 
Oscillations (BAO) \cite{BAO} placed tighter bounds 
on $w_{\rm DE}$. The cosmological constant has been 
overall consistent with the data, but the deviation of $w_{\rm DE}$ from $-1$ is also allowed \cite{Planckdark}.

{}From the theoretical side, there have been many attempts 
for constructing models of late-time cosmic acceleration
in General Relativity (GR) 
or modified gravitational 
theories \cite{review1,review2,review3,review4,review5,review6}. 
In GR, the representative dark energy models are 
quintessence \cite{quin1,quin2,quin3,quin4,quin5,quin6} 
and k-essence \cite{kes1,kes2,kes3}, in which 
the potential energy and the kinetic energy of a scalar field 
$\phi$ drives the acceleration, respectively. 
Provided that the ghost instability 
is absent, the dark energy equation of state 
of quintessence and k-essence
is in the range $w_{\rm DE}>-1$. 
In current observations, there has been no statistically 
strong observational evidence that the models 
with $w_{\rm DE}>-1$ are favored over the cosmological 
constant \cite{CDT}.

In modified gravitational theories, it is possible to realize 
$w_{\rm DE}$ less than $-1$, while satisfying conditions 
for the absence of ghosts and Laplacian instabilities. 
The models with $w_{\rm DE}<-1$ can reduce tensions of 
the Hubble constant $H_0$ between the CMB and the direct 
measurements of $H_0$ at low redshifts, so the 
best-fit model can be in the range $w_{\rm DE}<-1$ 
depending on the data analysis \cite{Planckdark,DHT17}.
In this sense, it is worthwhile to construct theoretically 
consistent dark energy models in modified gravitational theories and confront them with observations.

In the presence of a scalar field $\phi$ coupled to gravity, 
Horndeski theories \cite{Horndeski} are the most general 
scalar-tensor theories with second-order equations of 
motion \cite{Horn1,Horn2,Horn3}. 
There have been many attempts for constructing models 
of late-time cosmic acceleration in the framework 
of Horndeski theories, e.g., those in $f(R)$ 
gravity \cite{fR1,fR2,fR3,fR4}, 
Brans-Dicke theories \cite{BD1,BD2}, 
and Galileons \cite{Nicolis,Galileon1,Galileon2,GS10,DTGa1,DTGa2}.
At the background level, the dark energy equation of state 
in these models enters the region $w_{\rm DE}<-1$, 
so they can be distinguished from the $\Lambda$CDM model. 
Different modified gravity models also lead to 
different cosmic growth histories, so the observations of 
large-scale structure and weak lensing allow one to 
distinguish between the models.
For instance, the covariant Galileon is in strong tension 
with the data \cite{Nesseris,AppleLin,Totani,Barreira1,Neveu,Barreira2,Renk,Peirone} 
due to a very different structure 
formation pattern compared to the $\Lambda$CDM model \cite{FKT10}
and the large deviation of $w_{\rm DE}$
from $-1$ for tracker solutions \cite{DTGa1,DTGa2}.

One can perform a healthy extension of 
Horndeski theories in such a way that the number 
of propagating degrees of freedom 
is not increased \cite{Zuma,GLPV,Langlois1,Langlois2,Crisostomi16}. 
In GLPV theories \cite{GLPV}, for example, there are two 
additional Lagrangians beyond the domain of Horndeski theories. These beyond-Horndeski Lagrangians give rise to several distinguished features such as the mixing between the scalar and matter sound speeds \cite{Gergely,GLPV,KT14}, the breaking of the Vainshtein mechanism inside an astrophysical object \cite{Koba15,Sakstein,Mizuno,Babichev16}, and the appearance of a solid angle deficit singularity at the center of a compact body \cite{DKT15,DKT15b}. 
Even with these restrictions, it is possible to construct 
viable dark energy models in GLPV theories
without theoretical inconsistencies \cite{KTF15}.

The recent gravitational-wave (GW) event 
GW170817 \cite{GW170817}
from a neutron star merger together with the gamma-ray burst GRB 170817A \cite{Goldstein} significantly constrained the deviation of the propagation speed $c_t$ of GWs. In the natural unit where the speed of light is 1, the constraint on $c_t$ from the measurements of LIGO and Virgo 
is given by \cite{Abbott}
\be
-3 \times 10^{-15} \le c_t-1 \le 
7 \times 10^{-16}\,.
\label{ctbound}
\ee
Since the GWs have propagated over the cosmological 
distance from the redshift $z \simeq 0.009$ to us, 
the bound (\ref{ctbound}) can be applied to dark energy 
models in which $c_t$ is modified from 
the GR value ($c_t=1$) for $z \lesssim 10^{-2}$.

The modification from $c_t=1$ occurs in 
quartic- and quintic-order Horndeski theories
containing the dependence of the field kinetic energy 
$X=\nabla^{\mu}\phi \nabla_{\mu}\phi$ in the 
couplings $G_4$ and $G_5$ \cite{Horn2,DT11}. 
If we impose that $c_t^2$ is strictly equivalent to 1, 
the allowed Horndeski Lagrangians are up to the cubic interaction $G_3(\phi,X) 
\square \phi$ with the nonminimal 
coupling $G_4(\phi)R$, where $R$ 
is the Ricci scalar \cite{GW1,GW2,GW3,GW4,GW5,GW6,GW7} 
(see also Refs.~\cite{Lom1,Lom2}).
The $f(R)$ gravity and Brans-Dicke 
theories belong to this class, but the quartic- and quintic-order 
covariant Galileons lead to the deviation of $c_t^2$ from 1.

In GLPV theories, the existence of an additional contribution 
to the quartic-order Horndeski Lagrangian gives rise to a 
self-tuning cosmological model in which $c_t^2$ is exactly 
equivalent to 1 \cite{GW1,GW2}. 
In such theories, it was recently shown that the speed of GWs 
also equals to 1 on the background of an exact Schwarzschild-de Sitter solution \cite{Babi17}. 
This implies that the condition for imposing 
$c_t^2=1$ on the cosmological background can be 
sufficient to realize the same value in the vicinity of a strong gravitational source. 
In this case, the time delay does not occur when the GWs pass 
nearby massive objects. 
It is of interest to study whether the construction of  
viable dark energy models is possible in such self-tuning 
beyond-Horndeski theories. 

In this paper, we propose a simple dark energy model with a
self-accelerating de Sitter solution in the framework of GLPV theories. Since the quintic Lagrangian in GLPV theories 
leads to $c_t^2$ different from 1 \cite{building,Gergely,KT14,KaseIJ}, 
our analysis up to quartic-order
Lagrangians is a most general scheme allowing the exact 
value $c_t^2=1$ in the domain of GLPV theories. 
We will not restrict the models from the beginning and derive the background and perturbation equations of motion in 
a general way by exploiting useful dimensionless parameters 
introduced in Ref.~\cite{Bellini}.

Our dark energy model is a simple extension of covariant Galileons up to quartic order with the $a_2X^2$ term 
in the quadratic Lagrangian $L_2$. Since the linear Galileon 
term $a_1X$ is also present in $L_2$, the quadratic Lagrangian 
is similar to the ghost condensate model \cite{Arkani}. 
Our model also contains the beyond-Horndeski coupling 
$F_4$, which is constant. The self-tuning cosmological 
model mentioned above corresponds to the specific 
choice of $F_4$. 
We are primarily interested in this self-tuning theory, but 
we leave the constant $F_4$ arbitrary to discuss also how
the theories with $c_t^2 \neq 1$ can be constrained from 
the observational bound (\ref{ctbound}).

We would like to stress that the $a_2X^2$ term in $L_2$ is 
fundamentally important to realize the background solution different from 
the tracker arising in covariant Galileons. 
The tracker solution of covariant Galileons has the equation of state 
$w_{\rm DE}=-2$ during the matter era \cite{DTGa1,DTGa2}, 
which is ruled out from the joint data analysis of 
SN Ia, CMB, and BAO \cite{Nesseris}. 
In our model, the solutions can approach a self-accelerating 
de Sitter attractor ($X={\rm constant}$) before reaching
the tracker due to the presence of the term $a_2X^2$.
We will show that a variety of the dark energy equation of state 
including the entry to the region $w_{\rm DE}<-1$ can be realized, depending on the moment at which the term $a_2X^2$ dominates over cubic and quartic Galileon interactions.
Even if the dominant contribution to the field energy density 
in the early Universe corresponds to the quartic Galileon 
interaction, the conditions for the absence of ghosts and Laplacian instabilities can be consistently satisfied throughout the cosmic expansion history. 

We will also study the evolution of linear cosmological perturbations relevant to the observations of large-scale structure and compute the two quantities $\mu=G_{\rm eff}/G$ 
and $\Sigma=G_{\rm light}/G$ commonly used in the EFTCAMB 
code \cite{EFTCAMB1,EFTCAMB2,Pogosian}, 
where $G_{\rm eff}$ and $G_{\rm light}$ are gravitational 
couplings associated with the growth of matter perturbations 
and the light bending, respectively, 
with $G$ being the Newton constant. 
For the self-tuning model ($c_t^2=1$), we will show that 
the gravitational potentials $-\Psi$ and $\Phi$ are 
almost equivalent to each other,
in which case $\mu \simeq \Sigma > 1$. 
If the dominance of the term $a_2X^2$ occurs at a later cosmological epoch, $\mu$ and $\Sigma$ tend to deviate from 1.
Hence it is possible to distinguish our model from the $\Lambda$CDM model from both the background and cosmic growth histories. 

Nonlinear derivative interactions beyond Horndeski also 
have impacts on the gravitational law in regions of 
high density \cite{Koba15}.
We will study constraints on the deviation from 
Horndeski theories arising from the modification of 
gravitational potentials inside massive objects \cite{Sakstein,Mizuno,Babichev16,Jimenez,Sakstein1,Sakstein2,Jain15,Sakstein3,Cris2,Langlois17,Dima17}  
and show that our dark energy model with $c_t^2=1$ 
is well consistent with such bounds.

This paper is organized as follows. 
In Sec.~\ref{modelsec}, we revisit the no-ghost condition and the propagation speed of tensor perturbations in GLPV theories up to quartic-order.
In Sec.~\ref{backpersec}, we present the background and scalar perturbation 
equations without specifying the model and obtain general expressions of 
$\mu$ and $\Sigma$ under the quasi-static approximation on sub-horizon scales.
In Sec.~\ref{cosmosec}, we propose a concrete dark energy model in GLPV theories allowing for $c_t^2=1$. 
We study the background cosmological dynamics 
paying particular attention to the evolution of $w_{\rm DE}$ and investigate whether 
the conditions for the absence of ghosts and Laplacian instabilities are satisfied. 
In Sec.~\ref{vasec}, we discuss the evolution of linear 
cosmological perturbations and numerically compute the quantities $\mu$ and $\Sigma$ to confront the model 
with the observations of large-scale structure 
and weak lensing.
In Sec.~\ref{astrosec}, we study constraints on our model 
arising from corrections to gravitational potentials 
inside massive objects induced by the 
beyond-Horndeski nonlinear derivative interaction.
Sec.~\ref{consec} is devoted to conclusions.

\section{GLPV theories and the speed of 
gravitational waves}
\label{modelsec} 

The GLPV theories up to quartic order 
is given by the action \cite{GLPV}
\be
S=\int d^4 x \sqrt{-g}\,\sum^{4}_{i=2}L_i
+S_M\,,
\label{action}
\ee
where $g$ is a determinant of the metric $g_{\mu \nu}$
and $S_M$ is the matter action. 
We assume that the matter sector is minimally 
coupled to gravity.
The Lagrangians $L_{2,3,4}$ are given, 
respectively, by 
\ba
L_{2} &=&G_{2}(\phi,X)\,,  
\label{L2co} \\
L_{3} &=&G_{3}(\phi,X) \square \phi\,,  
\label{L3co} \\
L_{4} &=&G_{4}(\phi,X) R-2G_{4,X}(\phi,X)
\left[ (\square \phi )^{2}-\nabla ^{\mu} \nabla^{\nu}\phi 
\nabla_{\mu}\nabla_{\nu}\phi \right]  \notag \\
& &+F_{4}(\phi,X)\epsilon^{\mu\nu\rho\sigma}\epsilon_{\mu'\nu'\rho'\sigma} 
\nabla^{\mu'}\phi\nabla_{\mu}\phi
\nabla^{\nu'}\nabla_{\nu}\phi
\nabla^{\rho'}\nabla_{\rho}\phi\,, \label{L4}
\ea
where $G_{2,3,4}$ and $F_4$ depend on 
the scalar field $\phi$ and its kinetic energy 
$X\equiv \nabla^{\mu}\phi\nabla_{\mu}\phi$ 
with the partial derivative 
$G_{4,X}\equiv\partial G_4/\partial X$, 
and $R$, $G_{\mu\nu}$, $\epsilon_{\mu\nu\rho\sigma}$ 
are the Ricci scalar, the Einstein tensor, the totally
antisymmetric Levi-Civita tensor satisfying the normalization  
$\epsilon^{\mu\nu\alpha\beta}\epsilon_{\mu\nu\alpha\beta}=+4!$, respectively\footnote{The sign of 
$\epsilon^{\mu\nu\alpha\beta}$ is opposite to 
that used in Ref.~\cite{BGP} in the context of 
beyond generalized Proca theories.}. 
The last term of Eq.~(\ref{L4}) arises in theories 
beyond Horndeski. 
The quintic-order Lagrangian $L_5$ in GLPV theories 
gives rise to the speed $c_t$ of GWs 
different from 1 \cite{building,Gergely,KT14,KaseIJ}, 
so the action (\ref{action}) corresponds 
to the most general scalar-tensor theories allowing for $c_t=1$ 
in the framework of GLPV theories.

The action (\ref{action}) can be also written 
in terms of scalar quantities arising from the 
3+1 Arnowitt-Deser-Misner (ADM) decomposition 
of space-time \cite{ADM} with the foliation of constant-time 
hypersurfaces $\Sigma_t$ \cite{building}. 
The line element in the ADM formalism is given by 
$ds^2=g_{\mu \nu}dx^{\mu}dx^{\nu}= -N^{2}dt^{2}
+h_{ij}(dx^{i}+N^{i}dt)(dx^{j}+N^{j}dt)$, 
where $N$ is the lapse, $N^i$ is the shift, and 
$h_{ij}$ is the 3-dimensional spatial metric.
The extrinsic curvature and intrinsic curvature are 
defined, respectively, as 
$K_{\mu \nu}=h^{\lambda}_{\mu} n_{\nu;\lambda}$ 
and ${\cal R}_{\mu \nu}={}^{(3)}R_{\mu \nu}$, where
$n_{\mu}=(-N,0,0,0)$ is a normal vector orthogonal 
to $\Sigma_t$ and 
${}^{(3)}R_{\mu \nu}$ is the 3-dimensional Ricci tensor 
on $\Sigma_t$.
There are several scalar quantities constructed from 
$K_{\mu \nu}$ and ${\cal R}_{\mu \nu}$, as
\be
K \equiv g^{\mu \nu}K_{\mu \nu}\,,\qquad
{\cal S} \equiv K_{\mu \nu}K^{\mu \nu}\,,\qquad
{\cal R} \equiv g^{\mu \nu}{\cal R}_{\mu \nu}\,.
\ee
By choosing the so-called unitary gauge 
($\phi=\phi(t)$) for a time-like scalar field, 
the action (\ref{action}) is expressed in the form
$S=\int d^4x \sqrt{-g}\,L+S_m$, 
where \cite{building,GLPV}
\be
L=A_2(N,t)+A_3(N,t)K+A_4(N,t)(K^2-{\cal S})
+B_4(N,t){\cal R}\,,
\label{Lag}
\ee
with 
\be
A_2=G_2-XE_{3,\phi}\,,\qquad 
A_3=2\left|X\right|^{3/2} \left( E_{3,X}
+\frac{G_{4,\phi}}{X} \right)\,,\qquad
A_4=-G_4+2XG_{4,X}-X^2F_4\,,\qquad
B_4=G_4\,.
\label{A2re}
\ee
The auxiliary function $E_3(\phi, X)$ obeys 
\be
G_3=E_3+2XE_{3,X}\,.
\ee
For the time-dependent scalar field $\phi$, we have 
$X=-N^{-2}(d\phi/dt)^2<0$ and hence
$A_3=2(-X)^{3/2}E_{3,X}-2\sqrt{-X}G_{4,\phi}$. 
In the unitary gauge, the $\phi, X$ dependence in 
the functions $G_{2,3,4}$ and $F_4$ translates to the 
$N,t$ dependence in the functions $A_{2,3,4}$ and $B_4$. 
Thus, the Lagrangian (\ref{Lag}) depends on 
$N,t$ and $K,{\cal S},{\cal R}$.
The Horndeski theories satisfy the condition $F_4=0$, so that 
$A_4=-B_4+2XB_{4,X}$, while 
$F_4 \neq 0$ in GLPV theories.

We study the background dynamics of dark energy driven by 
the field $\phi$ as well as the evolution of linear cosmological 
perturbations for the perturbed line element:
\be
ds^2=-\left( 1+2\delta N \right)dt^2+2\partial_i \psi dt dx^i
+a^2(t) \left[ (1+2\zeta)\delta_{ij}+\gamma_{ij} 
\right] dx^i dx^j\,,
\label{permet}
\ee
where $a(t)$ is the time-dependent scale factor, 
$\delta N, \psi, \zeta$ are scalar metric perturbations, and 
$\gamma_{ij}$ are tensor perturbations. 
The scalar perturbation $E$, which appears as the form 
$\partial_i \partial_j E\,dx^i dx^j$ in Eq.~(\ref{permet}), 
is set to 0, so that the spatial component of a 
gauge-transformation vector $\xi^{\mu}$ is fixed. 
We also choose the unitary gauge in which the perturbation 
$\delta \phi$ of the scalar field $\phi$ vanishes, under which 
the temporal part of $\xi^{\mu}$ is fixed. 
In GLPV theories, there are no dynamical vector degrees of 
freedom, so we do not consider vector perturbations 
in our analysis.

The second-order tensor action derived by 
expanding (\ref{action}) up to quadratic order in 
$\gamma_{ij}$ is \cite{building,Gergely,KT14,KaseIJ}
\be
S_2^{(T)}=\int d^4 x\,a^3\, \frac{Q_{t}}{4} \delta^{ik} \delta^{jl}
\left( \dot{\gamma}_{ij} \dot{\gamma}_{kl}
-\frac{c_t^2}{a^2}\partial \gamma_{ij} 
\partial \gamma_{kl} \right)\,, 
\label{L2ten}
\ee
where a dot represents the derivative with 
respect to $t$, and 
\ba
Q_{t} &\equiv& L_{,\cal S}=-A_4\,,\label{qt}\\
c_{t}^2 &\equiv& \frac{L_{,\cal R}}{L_{,\cal S}}
=-\frac{B_4}{A_4}\,.
\label{qtctdef}
\ea
The condition for the absence of tensor ghosts corresponds 
to $Q_t>0$, i.e., $A_4<0$.
In GR we have $-A_4=B_4=1/(16\pi G)$, so 
$c_t^2$ is equivalent to 1. 
For the theories in which $-A_4$ is different from $B_4$, 
there is the deviation of $c_t^2$ from 1. 
The GLPV theories satisfying $c_t^2=1$ 
have the following relation
\be
F_4=\frac{2G_{4,X}}{X}\,,
\label{ctcon}
\ee
for $X \neq 0$.
For example, if we consider the case 
$G_4(X)=B_4(X)=1/(16\pi G)+b_4 X^2$ ($b_4$ is 
a constant), then the function $F_4=4b_4$ satisfies 
the condition (\ref{ctcon}). 
In Horndeski theories we have $F_4=0$, so
$G_4$ depends on $\phi$ alone. 
The existence of the additional Lagrangian containing 
the $F_4$ term in GLPV theories allows the possibility 
for realizing $c_t^2=1$ even for the function $G_4$ 
containing the $X$ dependence.

\section{Background and scalar perturbation equations}
\label{backpersec} 

In this section, we present the background and perturbation 
equations of motion in GLPV theories given by the action (\ref{action}). For the matter sector, 
we take into account a perfect fluid 
whose background density and pressure are given, 
respectively, by $\rho$ and $P$. 
The scalar perturbations of the matter energy-momentum tensor $T^{\mu}_{\nu}$ are expressed 
in the form 
\be
\delta T^{0}_{0}=-\delta \rho\,,\qquad
\delta T^{0}_{i}=\partial_i \delta q\,,\qquad
\delta T^{i}_{j}=\delta P \delta^i_j\,.
\ee
In the line element (\ref{permet}), 
there are also three scalar perturbations 
$\delta N, \psi, \zeta$ arising from the gravity sector.

We first revisit the general equations of motion without 
imposing the condition (\ref{ctcon}) and analytically 
estimate the quantities $\mu$ and $\Sigma$ under the 
condition that the deviation from Horndeski theories 
is not significant for the perturbations deep inside 
the Hubble radius.  
We then discuss how the constraint (\ref{ctbound}) of GWs 
puts restrictions on the values of $\mu$ and $\Sigma$.

\subsection{Background equations}

In the ADM language, the background equations of motion 
are expressed in the forms \cite{building}
\ba
& &
\bar{L}+L_{,N}-3H{\cal F}=\rho\,,
\label{back1d}\\
& &
\bar{L}-\dot{\cal F}-3H{\cal F}=-P\,,
\label{back2d}
\ea
where $H \equiv \dot{a}/a$ is the Hubble parameter, 
${\cal F} \equiv L_{,K}+2HL_{,{\cal S}}$,  and 
a bar represents quantities of the background. 
The matter sector obeys the continuity equation 
\be
\dot{\rho}+3H(\rho+P)=0\,.
\ee
In GLPV theories given by the Lagrangian (\ref{Lag}), 
Eqs.~(\ref{back1d}) and (\ref{back2d}) 
reduce, respectively, to 
\ba
& &
A_2-6H^2A_4+2\dot{\phi}^2 \left( A_{2,X}
+3HA_{3,X}+6H^2A_{4,X} \right)=\rho\,,
\label{back1}\\
& &
A_2-6H^2A_4-\dot{A}_3-4\dot{H}A_4 
-4H \dot{A}_4=-P\,,
\label{back2}
\ea
where we used $\bar{N}=1$, $\bar{K}=3H$, 
$\bar{\cal S}=3H^2$, and $\bar{\cal R}=0$. 
Since the function $B_4$ does not appear in 
Eqs.~(\ref{back1}) and (\ref{back2}), 
the two theories with same $A_{2,3,4}$ but 
with different $B_4$ cannot be distinguished from 
each other \cite{KT14}.
 In other words, Horndeski theories and 
GLPV theories with same functions $A_{2,3,4}$ 
lead to the same background cosmological dynamics.

\subsection{Scalar perturbation equations}

The second-order action of GLPV theories expanded 
up to quadratic order in scalar perturbations was 
already computed in 
Refs.~\cite{building,Gergely,KT14,KaseIJ}. 
To discuss the perturbation equations of motion for the Lagrangian 
(\ref{Lag}) containing the $N,t,K,{\cal S},{\cal R}$ dependence, it is 
convenient to define the following dimensionless 
quantities \cite{Bellini}
\be
\alpha_{\rm M} \equiv \frac{\dot{Q}_t}{HQ_t}\,,\qquad
\alpha_{\rm B} \equiv \frac{L_{,KN}+2HL_{,{\cal S}N}}{4HQ_t}\,,\qquad 
\alpha_{\rm K} \equiv \frac{2L_{,N}+L_{,NN}}{2H^2 Q_t}\,,\qquad
\alpha_{\rm H} \equiv \frac{L_{,\cal R}+L_{,N{\cal R}}}
{L_{,\cal S}}-1\,. 
\label{aldef}
\ee
The parameters $\alpha_{\rm M}, \alpha_{\rm B}, \alpha_{\rm K}$ 
correspond to the running of the gravitational constant, 
the kinetic mixing between the scalar field and gravity, and 
the kinetic term for the scalar, respectively.
The parameter $\alpha_{\rm H}$ characterizes 
the deviation from Horndeski theories. 
In quartic-order GLPV theories given by the Lagrangian 
(\ref{Lag}), it follows that 
\be
\alpha_{\rm H} = \frac{2XB_{4,X}-A_4-B_4}
{A_4}=\frac{X^2F_4}{A_4}\,,
\label{alH}
\ee
which does not vanish for $F_4 \neq 0$.

In Fourier space with the comoving wavenumber $k$, the perturbation equations of motion corresponding to Hamiltonian and momentum constraints, which are derived by varying the second-order action with respect to $\delta N$ and $\partial^2 \psi$, are given by \cite{Koyama}
\ba
& &
2H^2Q_t \left( \alpha_{\rm K}-12\alpha_{\rm B}-6 \right)
\delta N+4HQ_t (1+\alpha_{\rm B})\left( 3\dot{\zeta}+\frac{k^2}{a^2}\psi 
\right)+4Q_t \left( 1+\alpha_{\rm H} \right) 
\frac{k^2}{a^2}\zeta=\delta \rho\,,
\label{per1}\\
& & 
4HQ_t \left(1 +\alpha_{\rm B} \right)\delta N-4Q_t\dot{\zeta}
=-\delta q \,,
\label{per2}
\ea
respectively.
{}From the continuity equations 
${\delta T^{\mu}}_{0;\mu}=0$ and 
${\delta T^{\mu}}_{i;\mu}=0$, 
we obtain
\ba
& &
\dot{\delta \rho}+3H \left( \delta \rho+\delta P \right)
=-(\rho+P) \left( 3\dot{\zeta}+\frac{k^2}{a^2}\psi 
\right)+\frac{k^2}{a^2}\delta q\,,
\label{per3}\\
& &
\dot{\delta q}+3H \delta q=-\left( \rho+P \right) \delta N
-\delta P\,.
\label{per4}
\ea
Varying the second-order action of scalar perturbations 
with respect to $\zeta$ and using Eq.~(\ref{per4}), 
we obtain
\be
H\left( 1+\alpha_{\rm M} \right)\psi
+\dot{\psi}+\left( 1+\alpha_{\rm H} \right) \delta N
+c_t^2\zeta=0\,.
\label{per5}
\ee
The function $B_4$ appears in Eqs.~(\ref{per1}) and 
(\ref{per5}) through the terms $L_{,{\cal R}}=B_4$ and 
$L_{,N{\cal R}}=B_{4,N}$. 
Then, the two theories with same $A_{2,3,4}$ and with 
different $B_{4}$ can be distinguished from each 
other at the level of perturbations. 
By specifying the functional forms of $A_{2,3,4}$
and $B_{4}$ with a given equation of state $w=P/\rho$ 
and a matter sound speed squared 
$c_M^2=\delta P/\delta \rho$, we can solve 
Eqs.~(\ref{per1})-(\ref{per2}) and 
(\ref{per3})-(\ref{per5}) together with 
the background Eqs.~(\ref{back1})-(\ref{back2}) 
to determine $\delta N, \psi, \zeta, \delta \rho, \delta q$.

\subsection{Conditions for the absence of 
ghosts and Laplacian instabilities}

To study whether ghosts and Laplacian instabilities of 
scalar perturbations do not arise from the radiation 
era to today, we need to take into account both radiation and 
nonrelativistic matter in the action $S_M$.
The perfect fluids of radiation and nonrelativistic matter 
can be modeled by the purely k-essence 
Lagrangians (pressures)
$P_r(Y_r)=b_rY_r^2$ and 
$P_m(Y_m)=b_m(Y_m-Y_0)^2$, where 
$b_r,b_m,-Y_0$ are positive constants and 
$Y_r=\nabla^{\mu}\chi_r\nabla_{\mu}\chi_r$ and 
$Y_m=\nabla^{\mu}\chi_m\nabla_{\mu}\chi_m$ 
are the kinetic energies of two scalar fields 
$\chi_r$ and $\chi_m$ \cite{Scherrer,kinetic1,kinetic2,KT14,KaseIJ}. 
Since the background densities $\rho_i$ and the sound speed 
squares $c_i^2$ for $i=r,m$ are given, respectively, by 
$\rho_i=2Y_iP_{i,Y_i}-P_{i}$ and 
$c_i^2=P_{i,Y_i}/(P_{i,Y_i}-2\dot{\chi}_i^2P_{i,Y_iY_i})$, 
it follows that 
\ba
& &
w_r \equiv \frac{P_r}{\rho_r}=\frac13\,,\qquad\quad
c_r^2=\frac13\qquad \quad ({\rm for}~{\rm radiation})\,,\\
& &
w_m \equiv \frac{P_m}{\rho_m}=\frac{Y_m-Y_0}
{3Y_m+Y_0}\,,\qquad\quad
c_m^2=\frac{Y_m-Y_0}
{3Y_m-Y_0}\qquad \quad ({\rm for}~{\rm nonrelativistic~matter})\,.
\ea
Provided that $|(Y_m-Y_0)/Y_0| \ll 1$, both 
$w_m$ and $c_m^2$ are close to 0.

We also expand the matter action $S_M$ up to second 
order in scalar perturbations by considering the field perturbations $\delta \chi_r$ and $\delta \chi_m$ 
of radiation and nonrelativistic matter, respectively.
By using Eqs.~(\ref{per1}) and (\ref{per2}) to eliminate 
$\delta N$ and $\psi$ from the second-order scalar action 
of Eq.~(\ref{action}), the resulting action can be expressed 
in terms of the three perturbations 
$\delta \chi_r, \delta \chi_m,\zeta$ and their derivatives.
The no-ghost conditions associated with the perturbations 
$\delta \chi_r$ and $\delta \chi_m$ are trivially 
satisfied for $b_r>0$ and $b_m>0$, respectively.
The no-ghost condition for the dynamical scalar 
degree of  freedom $\zeta$ corresponds 
to \cite{building,Gergely,KT14,KaseIJ}
\be
Q_s \equiv \frac{Q_t (\alpha_{\rm K}+6\alpha_{\rm B}^2)}
{(1+\alpha_{\rm B})^2}>0\,.
\label{Qs}
\ee

Taking the small-scale limit in the second-order scalar action, 
we can also derive the three propagation speed squares
$c_s^2, \tilde{c}_r^2,\tilde{c}_m^2$ associated with the perturbations $\zeta,\delta \chi_r, \delta \chi_m$, respectively. 
In GLPV theories with $\alpha_{\rm H} \neq 0$, these propagation speeds are generally mixed 
with each other \cite{Gergely,GLPV,KT14}.
In the limit that $c_m^2 \to 0$, the matter sound speed 
squared $\tilde{c}_m^2$ is decoupled from others, such 
that $\tilde{c}_m^2=0$ \cite{Koyama}. 
The other two propagation speed squares are given, 
respectively, by \cite{Koyama,KTF15}
\ba
c_{s}^2 &=&
\frac12 \left[ c_r^2+c_{\rm H}^2-\beta_{\rm H}-
\sqrt{(c_r^2-c_{\rm H}^2+\beta_{\rm H})^2+
2c_r^2\alpha_{\rm H}\beta_r} \right]\,,\label{cs}\\
\tilde{c}_{r}^2 &=&
\frac12 \left[ c_r^2+c_{\rm H}^2-\beta_{\rm H}+
\sqrt{(c_r^2-c_{\rm H}^2+\beta_{\rm H})^2+
2c_r^2\alpha_{\rm H}\beta_r} \right]\,,\label{cr}
\ea
where 
\ba
& &
c_{\rm H}^2 =
\frac{2}{Q_s} \left[ \dot{\cal M}+H{\cal M}-c_t^2 Q_t
-\frac{ \rho_r+P_r+\rho_m+P_m}{4H^2(1+\alpha_{\rm B})^2} 
 \right]\,,
\qquad {\cal M}=\frac{Q_t (1+\alpha_{\rm H})}
{H(1+\alpha_{\rm B})}\,,\nonumber \\
& &
\beta_{\rm H}=\beta_r+\beta_m\,,\qquad 
\beta_r=\frac{\alpha_{\rm H} \left( \rho_r+P_r \right)}
{Q_s H^2(1+\alpha_{\rm B})^2}\,,
\qquad 
\beta_m=\frac{\alpha_{\rm H} \left( \rho_m+P_m \right)}
{Q_s H^2(1+\alpha_{\rm B})^2}\,.
\ea
In Horndeski theories we have $\alpha_{\rm H}=\beta_{\rm H}=0$, 
so Eqs.~(\ref{cs}) and (\ref{cr}) reduce to 
$c_s^2=c_{\rm H}^2$ and $\tilde{c}_r^2=c_r^2=1/3$, respectively. 
In GLPV theories with $|\alpha_{\rm H}| \ll 1$, it follows that 
\ba
c_s^2 &\simeq&  c_{\rm H}^2-\beta_{\rm H}
+\alpha_{\rm H} \beta_{r} 
\frac{c_r^2}{2(c_{\rm H}^2-c_r^2-\beta_{\rm H})}\,,\\
\tilde{c}_{r}^2 &\simeq&
c_{r}^2-\alpha_{\rm H}\beta_{r}
\frac{c_r^2}{2(c_{\rm H}^2-c_r^2-\beta_{\rm H})}\,.
\ea
Even when $|\alpha_{\rm H}| \ll 1$, the term 
$\beta_{\rm H}$ is not necessarily much smaller 
than 1 \cite{KT14}. 
Hence the deviation from 
Horndeski theories affects $c_s^2$ such that 
$c_s^2 \simeq c_{\rm H}^2-\beta_{\rm H}$, whereas 
$\tilde{c}_{r}^2$ is close to $c_r^2$.

\subsection{Sub-horizon perturbations}

We discuss the evolution of linear scalar perturbations 
for the modes relevant to the growth of large-scale structures.
We assume that the perfect fluid is described by 
nonrelativistic matter satisfying $P=0$ and 
$\delta P=0$, 
without taking into account the radiation. 
We introduce the gauge-invariant matter density contrast $\delta_m$ 
and gravitational potentials $\Psi, \Phi$, as
\be
\delta_m \equiv \delta-3V_m\,,\qquad 
\Psi \equiv \delta N +\dot{\psi}\,,\qquad
\Phi \equiv \zeta+H\psi\,,
\label{Psidef}
\ee
where $\delta \equiv \delta \rho/\rho$ and 
$V_m \equiv H \delta q/\rho$. 
Taking the time derivative of Eq.~(\ref{per3}) and using Eq.~(\ref{per4}), 
it follows that 
\be
\ddot{\delta}_m+2H\dot{\delta}_m+\frac{k^2}{a^2}\Psi
=-3 \left( \ddot{\cal B}+2H \dot{\cal B} \right)\,,
\label{delmeq}
\ee
where ${\cal B} \equiv \zeta+V_m$. 
The matter density contrast $\delta_m$ grows due to the 
gravitational instability associated with the potential $\Psi$. 
This relation is quantified by the modified 
Poisson equation 
\be
\frac{k^2}{a^2}\Psi=-4\pi \mu G\rho_m \delta_m\,,\quad 
{\rm with} \quad 
\mu=\frac{G_{\rm eff}}{G}\,,
\label{mudef}
\ee
where $G_{\rm eff}$ is the effective gravitational coupling 
generally different from the Newton constant $G$.
Introducing the gravitational slip parameter 
\be
\eta=-\frac{\Phi}{\Psi}\,,
\ee
the effective potential $\psi_{\rm eff}=\Phi-\Psi$ 
associated with the light bending in weak lensing observations obeys
\be
\frac{k^2}{a^2}\psi_{\rm eff}=
8\pi G\,\Sigma\rho_m \delta_m\,,\quad {\rm with} \quad 
\Sigma=\frac{1+\eta}{2} \mu\,.
\label{psieff}
\ee

On using the gravitational potentials $\Psi$ and $\Phi$, 
we can express Eq.~(\ref{per5}) as
\be
\left( 1+\alpha_{\rm H} \right)\Psi
+\left( 1+\alpha_{\rm M} \right)\Phi+
\left( c_t^2-1-\alpha_{\rm M} \right) \zeta
=\alpha_{\rm H} \dot{\psi}\,.
\label{quasi1}
\ee
Since we are interested in the evolution of perturbations 
for the modes deep inside the Hubble radius, we employ 
the approximation that the dominant contributions to 
perturbation equations are those containing 
$k^2/a^2$ and $\delta \rho$. This is called the 
quasi-static approximation \cite{quasi1,quasi2}, which is valid for 
the modes deep inside the sound horizon 
($c_s^2\,k^2/a^2 \gg H^2$) in a strict sense.
On using Eq.~(\ref{per2}), Eqs.~(\ref{per1}) and 
(\ref{per3}) reduce, respectively, to 
\ba
& &
\delta \rho \simeq 
4Q_t \frac{k^2}{a^2} \left[ \left(1+\alpha_{\rm B}\right)
\Phi-\left( \alpha_{\rm B}-\alpha_{\rm H} \right)\zeta 
\right] \,,\label{quasi2}\\
& &
\dot{\delta \rho}+3H\delta \rho-\frac{k^2}{a^2} 
\left[ \frac{\rho}{H} (\zeta-\Phi)+4Q_t \dot{\zeta} 
-4H Q_t (1+\alpha_{\rm B})(\Psi-\dot{\psi}) 
\right] \simeq 0\,.\label{per3d}
\ea
Substituting Eq.~(\ref{quasi2}) and its time derivative
into Eq.~(\ref{per3d}), we obtain
\ba
& &
\left( 1+\alpha_{\rm B} \right) \Psi+
\left[ \left( 1+\alpha_{\rm M}+h \right)
(1+\alpha_{\rm B})+\frac{\dot{\alpha}_{\rm B}}{H}
+\frac32 \tilde{\Omega}_m \right]\Phi \nonumber \\
& &
-\left[ \alpha_{\rm B} \left(1+\alpha_{\rm M}
+h\right)+h+\frac{\dot{\alpha}_{\rm B}}{H}
+\frac32 \tilde{\Omega}_m
-\alpha_{\rm H} \left(1+\alpha_{\rm M} \right)
-\frac{\dot{\alpha}_{\rm H}}{H} \right] \zeta 
\simeq -\frac{\alpha_{\rm H}}{H} \dot{\zeta}\,.
\label{quasi3}
\ea
where
\be
h \equiv \frac{\dot{H}}{H^2}\,,\qquad 
\tilde{\Omega}_m \equiv \frac{\rho_m}{6H^2Q_t}\,.
\ee
Since $\beta_r=0$ in the absence of radiation, the 
sound speed squared (\ref{cs}) reduces to 
$c_s^2=c_{\rm H}^2-\beta_m$, 
which is expressed in the form 
\be
c_s^2=\frac{2Q_t (1+\alpha_{\rm H})}
{Q_s(1+\alpha_{\rm B})} \left[ 1+\alpha_{\rm M}
-h-\frac{\dot{\alpha}_{\rm B}}{H(1+\alpha_{\rm B})}
+\frac{\dot{\alpha}_{\rm H}}{H(1+\alpha_{\rm H})}
-c_t^2\frac{1+\alpha_{\rm B}}{1+\alpha_{\rm H}}
-\frac{3(1+2\alpha_{\rm H})}
{2(1+\alpha_{\rm B})(1+\alpha_{\rm H})}
\tilde{\Omega}_m \right]\,.
\label{csex}
\ee

In GLPV theories with $\alpha_{\rm H} \neq 0$, 
there are two time derivatives on the r.h.s. of 
Eqs.~(\ref{quasi1}) and (\ref{quasi3}), so we 
cannot solve Eqs.~(\ref{quasi1}), (\ref{quasi2}), 
and (\ref{quasi3}) for $\Psi,\Phi,\zeta$.
If the parameter $|\alpha_{\rm H}|$ is not 
much smaller than 1, the oscillating mode of 
perturbations cannot be 
neglected relative to the mode induced by 
matter perturbations $\delta \rho$ in Eq.~(\ref{quasi2}). 
Indeed, the quasi-static approximation 
breaks down for the models in which 
$-A_4$ and $B_4$ are constants different from 
each other with $|\alpha_{\rm H}|={\cal O}(1)$ \cite{Koyama}. 
For the models with $|\alpha_{\rm H}| \ll 1$, 
the oscillating mode can be 
suppressed relative to the matter-induced 
mode \cite{KTF15}. 

In Horndeski theories or GLPV theories where 
$|\alpha_{\rm H}|$ is very much smaller than 1 
throughout the cosmic expansion history, we can 
derive the closed-form expressions of $\Psi,\Phi,\zeta$
by taking the limits $\alpha_{\rm H} \to 0$ and 
$\dot{\alpha}_{\rm H} \to 0$
in Eqs.~(\ref{quasi1}), (\ref{quasi2}), and (\ref{quasi3}). 
In doing so, we employ the approximation that 
$\delta_m \simeq \delta$ for sub-horizon perturbations 
and eliminate the $\dot{\alpha}_{\rm B}$ term in 
Eq.~(\ref{quasi3}) by using Eq.~(\ref{csex}). 
Then, the quantities $\mu$, $\eta$, and $\Sigma$ reduce, 
respectively, to 
\ba
\mu
&=& \frac{c_t^2}{16\pi GQ_t} 
\left[1+\frac{2Q_t \{c_t^2(1+\alpha_{\rm B})-1
-\alpha_{\rm M}\}^2}
{Q_s c_s^2 c_t^2(1+\alpha_{\rm B})^2} \right]\,,
\label{Gefff}\\
\eta &=&
\frac{Q_s c_s^2(1+\alpha_{\rm B})^2
+2Q_t \alpha_{\rm B}[c_t^2(1+\alpha_{\rm B})-1
-\alpha_{\rm M}]}
{Q_s c_s^2c_t^2(1+\alpha_{\rm B})^2+2Q_t 
[c_t^2(1+\alpha_{\rm B})-1
-\alpha_{\rm M}]^2}\,,
\label{etaf}\\
\Sigma &=& 
\frac{1+c_t^2}{32\pi GQ_t} 
\left[ 1+\frac{2Q_t \{ c_t^2(1+\alpha_{\rm B})
+\alpha_{\rm B}-1-\alpha_{\rm M}\}
\{ c_t^2(1+\alpha_{\rm B})-1-\alpha_{\rm M}\}}
{Q_s c_s^2(1+c_t^2)(1+\alpha_{\rm B})^2} \right]\,,
\label{Sigmaf}
\ea
which are also valid in full Horndeski theories 
including the quintic Lagrangian 
$L_5$ \cite{GLV,Tsuji15}. 
In GR with the functions $-A_4=B_4=1/(16\pi G)$ and 
$A_3=0$, we have $Q_t=1/(16\pi G)$, $c_t^2=1$, 
$\alpha_{\rm B}=\alpha_{\rm M}=0$, and hence 
$\mu=\eta=\Sigma=1$. 
In modified gravitational theories, the normalized 
effective gravitational coupling $\mu=G_{\rm eff}/G$ is 
composed of the ``tensor'' part $c_t^2/(16\pi GQ_t)$ and 
the ``scalar-matter interaction'' given by the second 
term in the square bracket of Eq.~(\ref{Gefff}). 
The scalar-matter interaction term is positive
under the stability conditions 
$Q_s>0, Q_t>0, c_s^2>0, c_t^2>0$. 
This reflects the fact that the fifth force induced by the scalar degree of freedom is attractive.

For the theories in which $c_t^2=1$, 
Eqs.~(\ref{Gefff})-(\ref{Sigmaf}) further simplify to 
\ba
\mu
&=& \frac{1}{16\pi GQ_t} 
\left[1+\frac{2Q_t (\alpha_{\rm B}-\alpha_{\rm M})^2}
{Q_s c_s^2 (1+\alpha_{\rm B})^2} \right]\,,
\label{Gefff2}\\
\eta &=&
\frac{Q_s c_s^2(1+\alpha_{\rm B})^2
+2Q_t\alpha_{\rm B}(\alpha_{\rm B}-\alpha_{\rm M})}
{Q_s c_s^2(1+\alpha_{\rm B})^2+2Q_t 
(\alpha_{\rm B}-\alpha_{\rm M})^2}\,, 
\label{eta2}\\
\Sigma &=& 
\frac{1}{16\pi GQ_t} 
\left[ 1+\frac{Q_t (2\alpha_{\rm B}-\alpha_{\rm M})
(\alpha_{\rm B}-\alpha_{\rm M})}
{Q_s c_s^2 (1+\alpha_{\rm B})^2} \right]\,,
\label{Sigmaf2}
\ea
respectively. 
Under the stability conditions $Q_s>0, Q_t>0, c_s^2>0$, 
the {\it necessary} condition for realizing 
$G_{\rm eff}$ smaller than 
$G$ corresponds to $1/(16\pi GQ_t)<1$, i.e., 
\be
Q_t>\frac{1}{16\pi G}\,.
\label{Qtcon}
\ee
As long as $\alpha_{\rm B} \neq \alpha_{\rm M}$, 
the scalar-matter interaction term
in Eq.~(\ref{Gefff2}) is positive, so 
the condition (\ref{Qtcon}) is not 
sufficient to realize $G_{\rm eff}<G$. 
For the opposite inequality to Eq.~(\ref{Qtcon}), 
$G_{\rm eff}$ is always larger than $G$. 
The quantity $\Sigma$ is greater
than $1/(16\pi GQ_t)$ for $(2\alpha_{\rm B}-\alpha_{\rm M})
(\alpha_{\rm B}-\alpha_{\rm M})>0$.

For the theories with $c_t^2 \neq 1$, the necessary condition 
for $\mu<1$ is modified to $Q_t>c_t^2/(16\pi G)$. 
In Ref.~\cite{Tsuji15}, the model with $c_t^2<1$ was proposed for realizing the small cosmic growth rate 
consistent with the observations of redshift-space distortions (RSDs) and CMB. 
This possibility is excluded after the GW170817 event, 
so we are left with only one parameter $Q_t$ 
for realizing $G_{\rm eff}<G$.

Modifications of the quantities $\mu$ 
and $\Sigma$ compared to those in GR lead to the 
different evolution of $\delta_m$ and $\psi_{\rm eff}$ through 
Eqs.~(\ref{delmeq}) and (\ref{psieff}). 
In Sec.~\ref{vasec}, we will study the dynamics of 
cosmological perturbations for a dark energy model 
with $\alpha_{\rm H} \neq 0$ without resorting to 
the approximation used for the derivation of 
Eqs.~(\ref{Gefff})-(\ref{Sigmaf}).

\section{Concrete dark energy model}
\label{cosmosec} 

In this section, we study the background cosmology for a concrete dark energy model given by the functions
\be
G_2=a_1X+a_2X^2\,,\qquad G_3=3a_3X\,,\qquad 
G_4=\frac{1}{16\pi G}+b_4X^2\,,\qquad 
F_4=3b_4-a_4\,,
\label{G234}
\ee
where $a_{1,2,3,4}$ and $b_4$ are constants. 
The model contains the Galileon 
interactions \cite{Galileon1} with 
the additional $a_2X^2$ term. 
For $a_1>0$ and $a_2>0$, the function 
$G_2=a_1X+a_2X^2$ corresponds to the ghost condensate 
scenario \cite{Arkani} in which the cosmic acceleration is realized by the de Sitter fixed point satisfying 
$G_{2,X}=0$, i.e., $X=-a_1/(2a_2)$. 
The existence of cubic and quartic Galileon interactions modifies the way how the solutions approach 
the de Sitter fixed point.
Alternatively, we can consider a quintessence scenario 
in which the  cosmic acceleration is driven 
by a scalar potential $V(\phi)$ \cite{KTF15}, 
i.e., $G_2=-X/2-V(\phi)$. It is also possible to generalize 
the Einstein-Hilbert term in $G_4$ to a nonminimal 
coupling of the form $F(\phi)/(16\pi G)$.
In this paper, we are interested in the self-accelerating solution 
approaching a constant value $X$, 
so the explicit $\phi$ dependence is not included in the 
functions $G_{2,3,4}$ and $F_4$.

On using the correspondence (\ref{A2re}), the model given 
by the functions (\ref{G234}) is equivalent to the Lagrangian 
(\ref{Lag}) with 
\be
A_2=a_1X+a_2X^2\,,\qquad A_3=2a_3|X|^{3/2}\,,\qquad 
A_4=-\frac{1}{16\pi G}+a_4X^2\,,\qquad 
B_4=\frac{1}{16\pi G}+b_4X^2\,,
\label{model}
\ee
and $E_3=a_3X$. The tensor propagation speed squared 
$c_t^2$ and the parameter $\alpha_{\rm H}$ are given,  
respectively, by 
\be
c_t^2=\frac{1+16\pi G\,sa_4X^2}
{1-16\pi G\,a_4X^2}\,,\qquad 
\alpha_{\rm H}=\frac{16\pi Ga_4X^2}
{1-16\pi Ga_4X^2}(1-3s)\,,
\label{aH}
\ee
where 
\be
s \equiv \frac{b_4}{a_4}\,.
\ee
For the theories with $s=-1$, $c_t^2$ is equivalent to 1, 
but $\alpha_{\rm H}$ is different from 0. 
The covariant Galileon in Horndeski theories corresponds to 
$s=1/3$, in which case $c_t^2$ deviates from 1.

\subsection{Dynamical system}

To discuss the background cosmological dynamics, we take into 
account radiation (density $\rho_r$ and pressure $P_r=\rho_r/3$) 
and nonrelativistic matter (density $\rho_m$ and pressure $P_m=0$). 
Substituting the functions (\ref{model}) into 
Eqs.~(\ref{back1}) and (\ref{back2}), the background equations 
of motion are 
\ba
& &
3H^2=8\pi G \left( \rho_{\rm DE}+\rho_r
+\rho_m \right)\,,
\label{back1f}\\
& &
2\dot{H}+3H^2 
=-8\pi G \left( P_{\rm DE}+P_r \right)\,,
\label{back2f}
\ea
where 
\ba
& &
\rho_{\rm DE}=-a_1 \dot{\phi}^2+3a_2 \dot{\phi}^4
+18a_3H \dot{\phi}^3+30a_4H^2 \dot{\phi}^4\,,\\
& &
P_{\rm DE}=-a_1 \dot{\phi}^2+a_2  \dot{\phi}^4
-6a_3 \dot{\phi}^2 \ddot{\phi}
-2a_4 \dot{\phi}^3 \left[ 8H \ddot{\phi}
+\dot{\phi}(2\dot{H}+3H^2) \right]\,.
\ea
We define the following dimensionless quantities:
\be
x_1 \equiv -\frac{8\pi G a_1\dot{\phi}^2}{3H^2}\,,
\qquad 
x_2 \equiv \frac{8\pi G a_2\dot{\phi}^4}{H^2}\,,
\qquad 
x_3 \equiv \frac{48\pi Ga_3\dot{\phi}^3}{H}\,,
\qquad 
x_4 \equiv 80\pi Ga_4 \dot{\phi}^4\,,
\qquad 
\Omega_r \equiv \frac{8\pi G\rho_r}{3H^2}\,,
\ee
which correspond to density 
parameters arising from the couplings 
$a_1X, a_2 X^2, 2a_3 |X|^{3/2}, a_4X^2$, 
and radiation, respectively. 
It is convenient to use 
$x_1, x_2, x_3, x_4$ rather than the coupling constants 
$a_1, a_2, a_3, a_4$ themselves, since we can easily 
make a comparison between different 
energy densities arising from 
different couplings.
The standard minimally coupled scalar field 
without the potential ($a_1=-1/2$) is recovered by 
taking the limits $x_1 \to 4\pi G \dot{\phi}^2/(3H^2)$, 
$x_2 \to 0$, $x_3 \to 0$, $x_4 \to 0$.

{}From Eq.~(\ref{back1f}), 
it follows that 
\be
\Omega_m \equiv \frac{8\pi G\rho_m}{3H^2}
=1-\Omega_{\rm DE}-\Omega_r\,,\qquad 
\Omega_{\rm DE} \equiv x_1+x_2+x_3+x_4\,.
\label{Omem}
\ee
The quantities $x_{1,2,3,4}$ and $\Omega_r$ obey the 
differential equations 
\ba
x_1' &=& 2x_1 \left( \epsilon_{\phi}-h \right)\,,
\label{x1eq} \\
x_2' &=& 2x_2 \left( 2\epsilon_{\phi}-h \right)\,,\\
x_3' &=& x_3 \left( 3\epsilon_{\phi}-h \right)\,,\\
x_4' &=& 4x_4 \epsilon_{\phi}\,,
\label{x4eq} \\
\Omega_r' &=& -2\Omega_r \left( 2+h \right)\,,
\label{Omereq}
\ea
where a prime represents a derivative with respect to 
${\cal N}=\ln a$, and 
\ba
\epsilon_{\phi} &\equiv& \frac{\ddot{\phi}}{H \dot{\phi}}
=-\frac{1}{q_s}\left[20(3x_1+2x_2)-5x_3(3x_1+x_2+\Omega_r-3)
-x_4(36x_1+16x_2+3x_3+8\Omega_r)\right]\,,
\label{epphi}\\
h &\equiv&\frac{\dot{H}}{H^2}=
-\frac{1}{q_s}
[10(3x_1+x_2+\Omega_r+3)(x_1+2x_2)
+10x_3(6x_1+3x_2+\Omega_r+3)+15x_3^2 \nonumber \\
& &\qquad\qquad\quad+x_4(78x_1+32x_2+30x_3+12\Omega_r+36)
+12x_4^2]\,,
\label{dh}
\ea
with 
\be
q_s \equiv 20(x_1+2x_2+x_3)
+4x_4 (6-x_1-2x_2+3x_3)
+5x_3^2+8x_4^2\,.
\label{qs}
\ee
The dark energy equation of state is given by 
\be
w_{\rm DE} \equiv \frac{P_{\rm DE}}{\rho_{\rm DE}}
=\frac{5(3x_1+x_2-\epsilon_{\phi}x_3)-(3+8\epsilon_{\phi}+2h)x_4}
{15(x_1+x_2+x_3+x_4)}\,.
\label{wde}
\ee

\subsection{de Sitter fixed point}

For the dynamical system (\ref{x1eq})-(\ref{Omereq}), 
there exists a de Sitter fixed point satisfying 
$\epsilon_{\phi}=0$, $h=0$, and $\Omega_r=0$, i.e., 
both $\dot{\phi}$ and $H$ are constants.
{}From Eqs.~(\ref{epphi}) and (\ref{dh}), we obtain the  
two relations among four constants 
$x_1,x_2,x_3,x_4$, as
\be
x_4=1-x_1-x_2-x_3\,,\qquad 
x_3=-\frac13 \left( 18x_1+8x_2+12 \right)\,,
\label{dSre}
\ee
on the de Sitter solution. {}From Eq.~(\ref{Omem}), 
it follows that $\Omega_{\rm DE}=1$ and $\Omega_m=0$.
Substituting the two relations (\ref{dSre}) with $\epsilon_{\phi}=0$ and $h=0$ into Eq.~(\ref{wde}), we have 
$w_{\rm DE}=-1$ as expected.

The stability of the de Sitter solution is known by considering 
the homogenous perturbations $\delta \epsilon_{\phi}, \delta h, \delta\Omega_r$
of the quantities $\epsilon_{\phi}, h, \Omega_r$, respectively. 
Since $\epsilon_{\phi}=h=0$ on the de Sitter fixed point, the perturbation $\delta x_1$ of the variable $x_1$ obeys $\delta x_1'=2x_1(\delta \epsilon_{\phi}-\delta h)$. 
On using the fact that the similar properties hold for 
$\delta x_2, \delta x_3, \delta x_4, \delta \Omega_r$
and taking the ${\cal N}$ derivative of 
Eqs.~(\ref{epphi}) and (\ref{dh}), it follows that the perturbations 
${}^{t}{\bm X}=(\delta \epsilon_{\phi}, \delta h, \delta\Omega_r)$ 
around the de Sitter fixed point obey
\be
{\bm X}'=A{\bm X}\,,
\ee
where $A$ is the $3 \times 3$ matrix whose nonvanishing components are 
given by 
\ba
&&A_{11}=-3\,,\qquad A_{22}=-3\,,\qquad A_{33}=-4\,,\nonumber \\
&&A_{13}=\frac{18(x_1+2)}{3x_1(5x_2-6)+6x_2+4x_2^2-36}\,,\qquad 
A_{23}=-\frac{6(3x_1+4x_2+6)}{3x_1(5x_2-6)+6x_2+4x_2^2-36}\,.
\ea
Since the eigenvalues of $A$ are $-3, -3,-4$, the de Sitter fixed point  
is always stable. This means that the solutions finally approach
the de Sitter attractor, independent of the initial conditions 
of $x_1,x_2,x_3,x_4$.

\subsection{Dark energy dynamics}

We study the dynamics of dark energy from the 
radiation-dominated epoch to today. 
As in the ghost condensate model \cite{Arkani}, we will 
focus on the case in which $x_1$ is negative, while 
$x_2, x_3,x_4$ are positive.

In the radiation and deep matter eras, 
let us consider the situation in which the quantities 
$x_3$ and $x_4$ dominate over $|x_1|$ and $x_2$. 
We also use the fact that $x_3$ and $x_4$ are much 
smaller than 1 in these epochs, by reflecting 
that $\Omega_{\rm DE} \ll 1$.
Then, the quantities (\ref{epphi}) and (\ref{dh}) 
reduce, respectively, to 
\be
\epsilon_{\phi} \simeq \frac{5(\Omega_r-3)x_3
+8\Omega_r x_4}{4(5x_3+6x_4)}\,,\qquad 
h \simeq -\frac{1}{2} \left( 3+\Omega_r 
\right)\,.
\label{epphih}
\ee

\begin{figure}[h]
\begin{center}
\includegraphics[height=3.3in,width=3.4in]{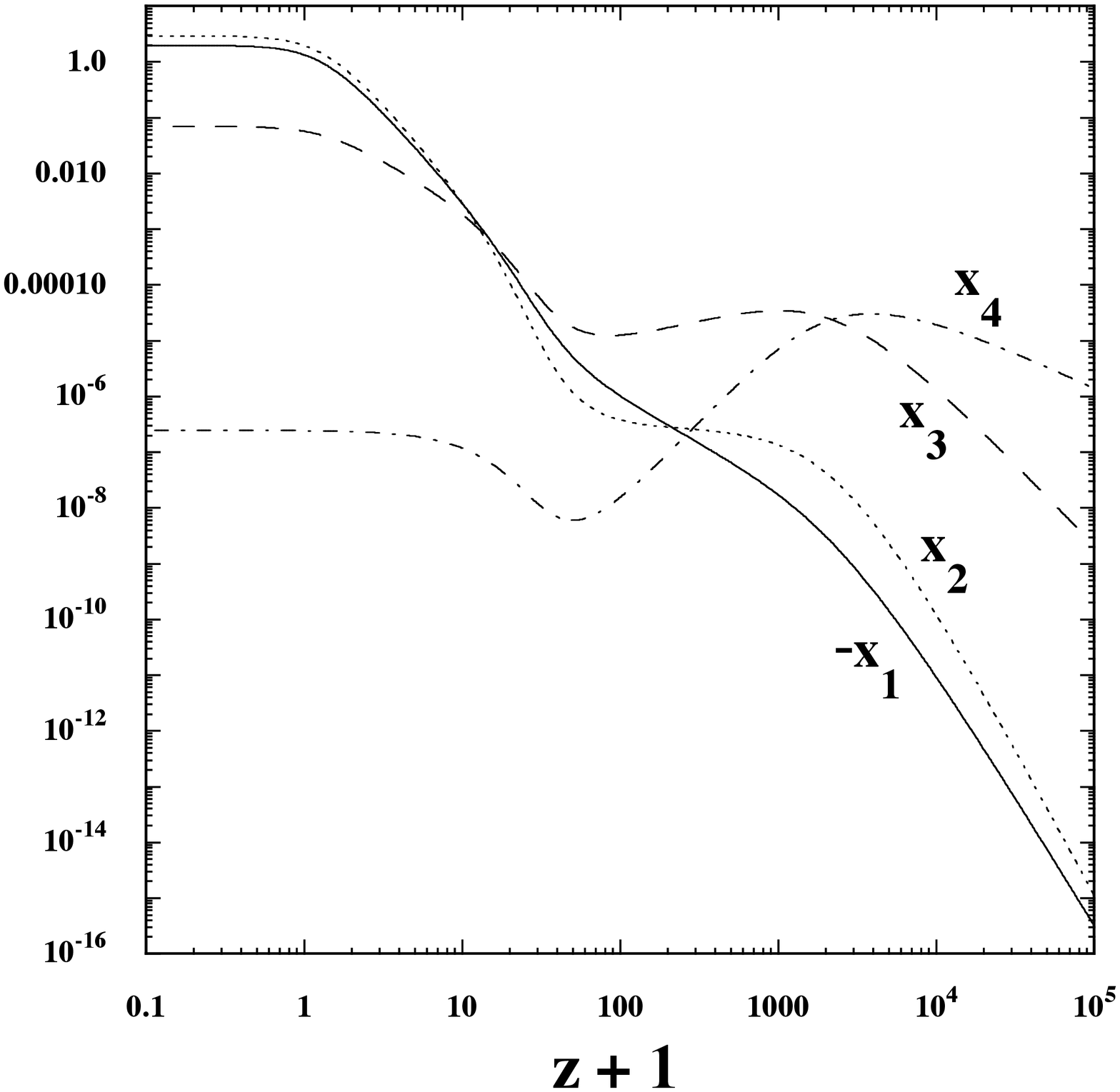}
\includegraphics[height=3.3in,width=3.4in]{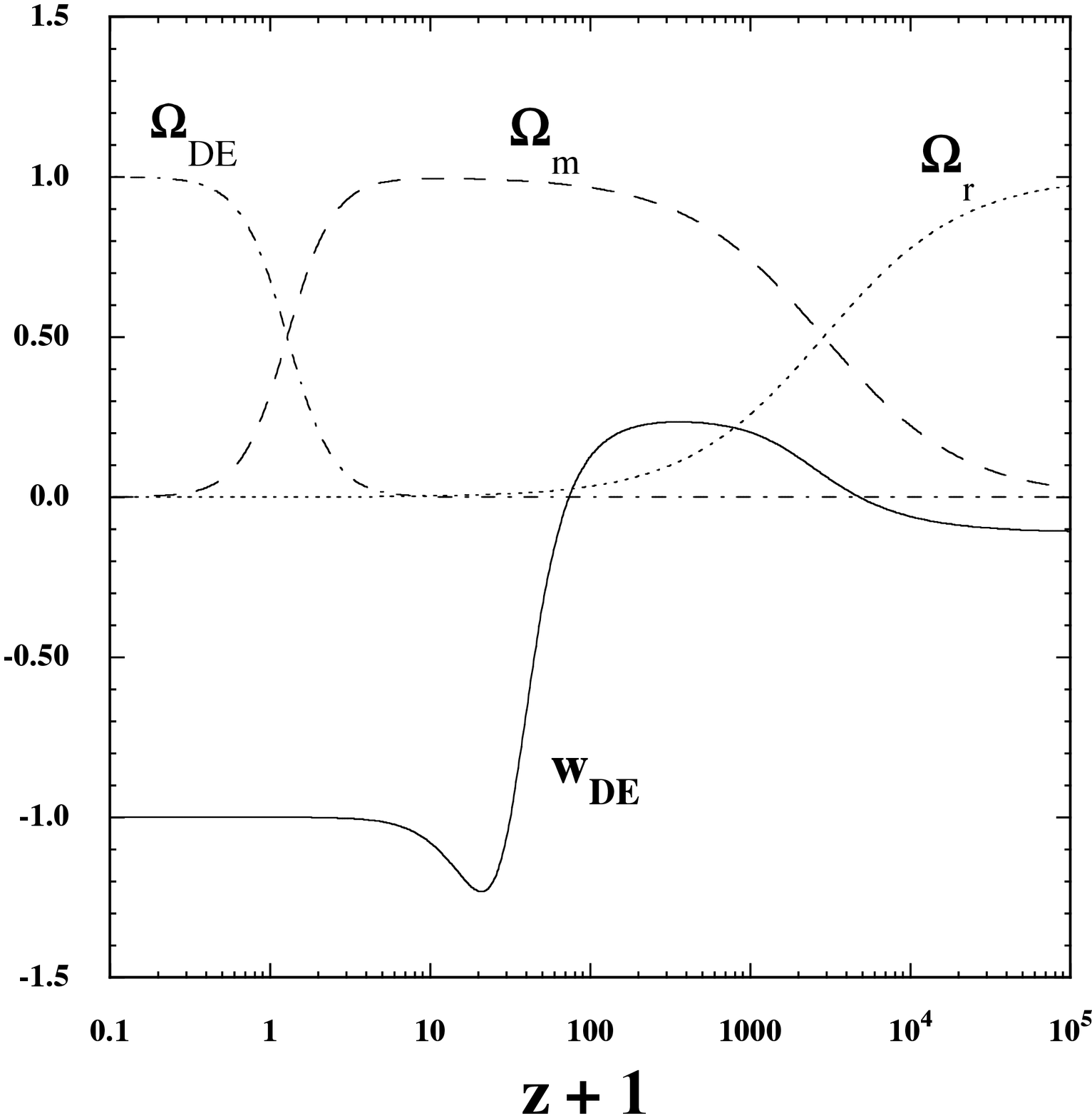}
\end{center}
\caption{\label{fig1}
Evolution of $-x_1, x_2, x_3, x_4$ (left) and 
$w_{\rm DE}, \Omega_{\rm DE}, \Omega_r, \Omega_m$ 
(right) versus $z+1$ (where $z=1/a-1$) with 
the initial conditions $x_1=-1.0 \times 10^{-16}$, 
$x_2=3.0 \times 10^{-16}$, $x_3=1.0 \times 10^{-9}$, 
$x_4=1.0 \times 10^{-6}$, and $\Omega_r=0.975$ 
at the redshift $z=1.3 \times 10^5$. 
We identify the present epoch according to 
$\Omega_{\rm DE}(z=0)=0.68$.
}
\end{figure}

\begin{figure}[h]
\begin{center}
\includegraphics[height=3.3in,width=3.4in]{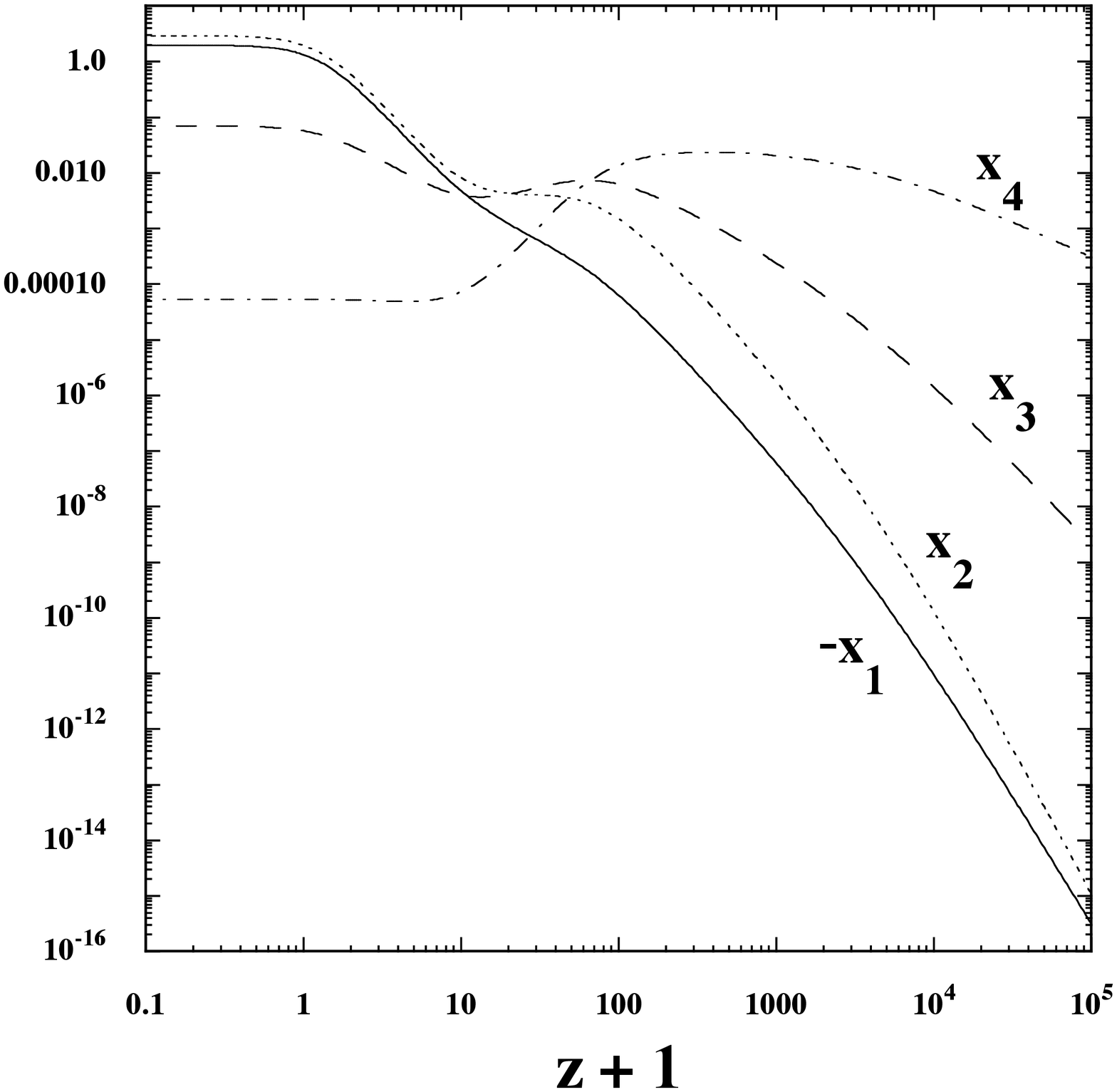}
\includegraphics[height=3.3in,width=3.4in]{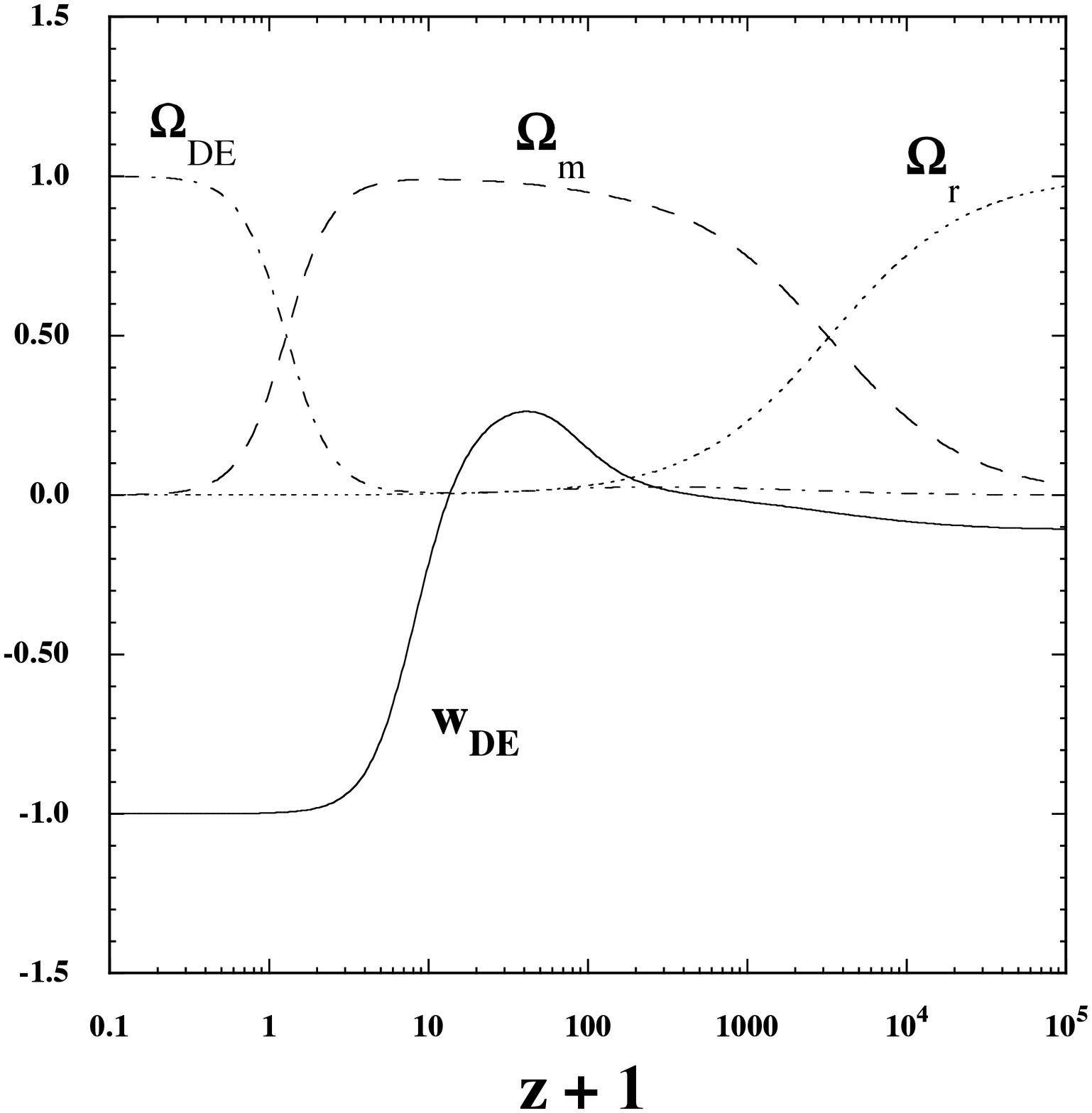}
\end{center}
\caption{\label{fig2}
The same as Fig.~\ref{fig1}, but with the different initial 
condition $x_4=2.2 \times 10^{-4}$
at $z=1.3 \times 10^5$. 
}
\end{figure}

If $x_4 \gg \{|x_1|, x_2, x_3\}$, then we have 
$\epsilon_{\phi} \simeq \Omega_r/3$. 
In this case, $\epsilon_{\phi} \simeq 1/3$ and 
$h \simeq -2$ during the radiation domination 
($\Omega_r \simeq 1$), so Eqs.~(\ref{x1eq})-(\ref{x4eq}) 
can be solved to give 
$|x_1| \propto a^{14/3}$, 
$x_2 \propto a^{16/3}$, 
$x_3 \propto a^{3}$, and 
$x_4 \propto a^{4/3}$. 
This behavior of $x_1, x_2, x_3, x_4$ can 
be confirmed in the numerical simulations of 
Figs.~\ref{fig1} and \ref{fig2} 
at the redshift $z \gtrsim 3000$.
{}From Eq.~(\ref{wde}), the dark energy equation 
of state in the regime $x_4 \gg \{|x_1|, x_2, x_3\}$ is given by 
\be
w_{\rm DE} \simeq -\frac{1}{9}\Omega_r\,.
\label{wdeap1}
\ee
In the numerical simulation of Figs.~\ref{fig1} and \ref{fig2}, 
$x_4$ dominates over $|x_1|, x_2, x_3$ during the radiation 
domination, so that $w_{\rm DE} \simeq -1/9$.
If the condition $x_3 \gg \{|x_1|, x_2, x_4\}$ is satisfied 
in the radiation and deep matter eras, 
the dark energy equation of state (\ref{wde}) yields 
\be
w_{\rm DE}=\frac{1}{4}-\frac{1}{12}\Omega_r\,.
\label{wdeap2}
\ee
In Fig.~\ref{fig1}, the dominance of $x_3$ over $x_4, |x_1|, x_2$ 
starts to occur right after the onset of the matter-dominated epoch, 
so $w_{\rm DE}$ temporally approaches the value $1/4$.
In Fig.~\ref{fig2}, $x_4$ dominates over $|x_1|, x_2, x_3$ by 
the redshift $z \sim 100$, so the evolution of $w_{\rm DE}$ 
for $z \gtrsim 100$ is approximately given by  
Eq.~(\ref{wdeap1}). 
In the limit that $\Omega_r \to 0$, Eq.~(\ref{wdeap1}) 
reduces to $0$, so the term $x_4$ arising from the quartic 
Galileon works as dark matter during the matter dominance. 
In Fig.~\ref{fig2}, we can confirm that the variable $x_4$ stays 
nearly constant around $0.01$ for 
$100 \lesssim z \lesssim 1000$.

After $x_3$ dominates over $|x_1|, x_2, x_4$ during the matter era, 
the quantities given in Eq.~(\ref{epphih}) reduce to 
$\epsilon_{\phi} \simeq -3/4$ and 
$h \simeq -3/2$, so the solutions to 
Eqs.~(\ref{x1eq})-(\ref{x4eq}) are given, respectively, by 
$|x_1| \propto a^{3/2}, x_2 \propto a^0, 
x_3 \propto a^{-3/4}$, and $x_4 \propto a^{-3}$. 
Then, the quantities 
$|x_1|$ and $x_2$ eventually catch up with 
$x_3$ and $x_4$. 
After $|x_1|$ and $x_2$ grows to the order of 1, 
the energy density associated with the Lagrangian 
$G_2=a_1X+a_2X^2$ becomes the main source for the 
late-time cosmic acceleration. There are also contributions 
to the dark energy density arising from the cubic and quartic 
Galileon terms. Since $x_4$ decreases more significantly than 
$x_3$ during the matter era, today's value of $x_4$ is 
typically much smaller than $x_3$ and $|x_1|, x_2$, see 
Figs.~\ref{fig1} and \ref{fig2}. 
Demanding that the maximum value of $x_4$ reached during the 
matter era does not exceed the order of 0.1, today's value of $x_4$ 
is constrained to be $x_4^{(0)} \lesssim 10^{-3}$.

In Fig.~\ref{fig1}, the dark energy equation of state enters 
the region $w_{\rm DE}<-1$ first, takes a minimum, 
and then finally approaches the de Sitter value  $-1$. 
In Fig.~\ref{fig2}, the solution reaches the de Sitter 
attractor without entering the region $w_{\rm DE}<-1$.
This difference comes from how much extent the cubic Galileon 
term $x_3$ contributes to the dark energy density at late times. 
If there is a long period in which $x_3$ dominates over 
$x_2$ and $|x_1|$ as in the case of Fig.~\ref{fig1}, the solutions 
enter the region $w_{\rm DE}<-1$. 
This reflects the fact that the tracker present for $x_2 \to 0$ 
has the equation of state smaller than $-1$ \cite{DTGa1,DTGa2}.
However, if this period is short as in the case of Fig.~\ref{fig2}, 
the k-essence terms $x_2$ and $|x_1|$ start to dominate over $x_3$ 
before the entry to the region
$w_{\rm DE}<-1$.

\begin{figure}[h]
\begin{center}
\includegraphics[height=3.2in,width=3.4in]{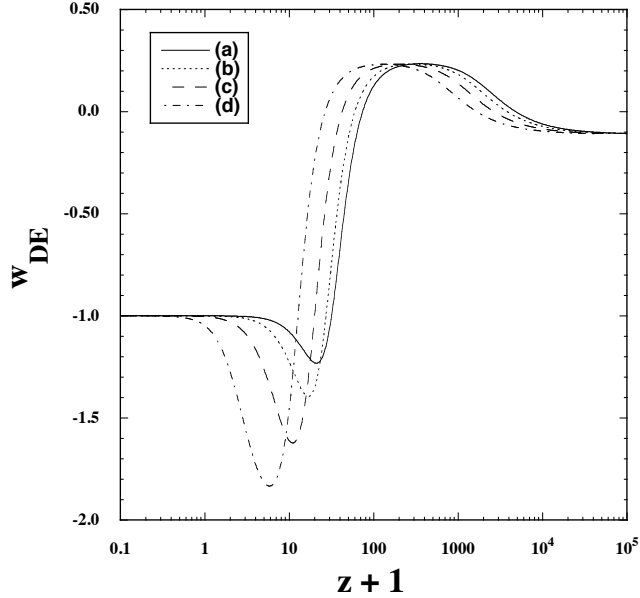}
\end{center}
\caption{\label{fig3}
Evolution of the dark energy equation of state $w_{\rm DE}$ 
versus $z+1$ for four different initial values of $x_2$, 
while the initial conditions of $x_1, x_3, x_4, \Omega_r$ 
are the same as those used in Fig.~\ref{fig1}. 
Each plot corresponds to 
(a) $x_2=3.0 \times 10^{-16}$ at $z=1.3 \times 10^5$, 
(b) $x_2=1.5 \times 10^{-16}$ at $z=1.0 \times 10^5$, 
(c) $x_2=5.0 \times 10^{-17}$ at $z=7.1 \times 10^4$, and 
(d) $x_2=1.0 \times 10^{-17}$ at $z=3.3 \times 10^4$.
}
\end{figure}

The tracker solution, which exists for $x_2 \to 0$, 
satisfies the relation \cite{DTGa1,DTGa2}
\be
H \dot{\phi}={\rm constant} \qquad i.e., 
\qquad 
\epsilon_{\phi}=-h\,.
\label{tra1}
\ee
On using Eqs.~(\ref{epphi}) and (\ref{dh}) and taking the 
limit $x_2 \to 0$,  this condition translates to 
\be
10x_1+5x_3+4x_4=0\,.
\label{tra2}
\ee
Substituting the relations (\ref{tra1}) and (\ref{tra2})
into Eq.~(\ref{wde}), it follows that 
\be
w_{\rm DE}=-1+\frac{2}{3}h\,,
\ee
which is equivalent to $-2$ during 
the matter dominance ($h\simeq-3/2$).

Now, we are considering the theories with $x_2 \neq 0$, so 
the approach to the tracker is prevented by the $a_2X^2$ term. 
In Fig.~\ref{fig3}, we plot the evolution 
of $w_{\rm DE}$ for several different initial values of $x_2$ 
with same initial conditions of $x_1, x_3, x_4, \Omega_r$ 
as those used in Fig.~\ref{fig1}.
In the limit that $x_2 \to 0$, the solutions 
approach the tracker equation of state 
$w_{\rm DE}=-2$ during the matter era.
This tracker equation of state is in tension with the 
joint data analysis of SNIa, CMB, and BAO \cite{Nesseris}.
In the present case, however, the $a_2X^2$ term 
leads to $w_{\rm DE}$ larger than $-2$.
For smaller $x_2$, the solutions enter the stage in which 
the terms $x_2$ and $|x_1|$ dominate over $x_3$ and 
$x_4$ later, so the 
minimum values of $w_{\rm DE}$ tend to be smaller. 
This property can be confirmed in the numerical 
simulation of Fig.~\ref{fig3}.
In case (a) of Fig.~\ref{fig3}, which corresponds to 
the initial conditions of Fig.~\ref{fig1}, the 
dark energy equation of state takes a minimum 
$w_{\rm DE} \simeq -1.2$ around the 
redshift $z \simeq 20$ and then approaches 
the asymptotic value $-1$. 
In case (d), the minimum value 
$w_{\rm DE} \simeq -1.83$ is reached around $z \simeq 5$. 
It will be of interest to study how the joint data analysis 
of SNIa, CMB, and BAO place bounds on today's 
values of of $x_1,x_2,x_3,x_4$.

\begin{figure}[h]
\begin{center}
\includegraphics[height=3.2in,width=3.4in]{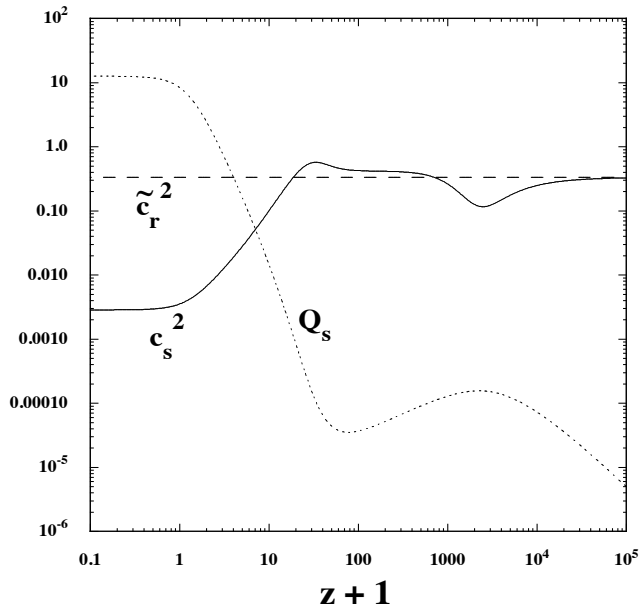}
\end{center}
\caption{\label{fig4}
Evolution of the quantities $Q_s$, $c_s^2$, and 
$\tilde{c}_r^2$ versus $z+1$ 
for the model $s=-1$ (i.e., $c_t^2=1$). 
The initial conditions of $x_1,x_2,x_3,x_4$ and 
$\Omega_r$ are the same 
as those used in Fig.~\ref{fig1}.
}
\end{figure}

\subsection{Stability conditions}

We study whether the conditions for the absence of ghosts 
and Laplacian instabilities are satisfied for the background 
cosmology discussed above. 
The quantities (\ref{qt}) and (\ref{qtctdef}) reduce, 
respectively, to 
\be
Q_t=\frac{5-x_4}{80\pi G}\,,\qquad 
c_t^2=\frac{5+sx_4}{5-x_4}\,.
\label{ctvalue}
\ee
The ghost and Laplacian instabilities of tensor perturbations 
are absent under the conditions
\ba
& &
x_4<5\,,\label{tennoghost}\\
& &
5+sx_4>0\,.\label{tennoghost2}
\ea
Since $x_4$ does not exceed the order of 1 for the realization 
of viable cosmology, the two conditions (\ref{tennoghost}) 
and (\ref{tennoghost2}) are well satisfied for $|s|={\cal O}(1)$. 
The recent detection of gravitational waves 
from a binary neutron star merger placed the bound 
(\ref{ctbound}) for the redshift $z<0.009$. 
On using Eq.~(\ref{ctvalue}), today's value of $x_4$, 
denoted as $x_4^{(0)}$, is 
constrained to be 
\be
\left| (s+1)x_4^{(0)} \right| \lesssim 10^{-14}\,.
\label{x4bound}
\ee
For the theories with $s=-1$, $c_t^2$ is equivalent to 1. 
In this case, the bound (\ref{x4bound}) is trivially satisfied 
irrespective of the value of $x_4^{(0)}$. 
However, for the theories in which $s$ is different from 
$-1$, $x_4^{(0)}$ is constrained to be smaller than 
the order of $10^{-14}$ for $s={\cal O}(1)$. 
This is the case for covariant Galileons in Horndeski theories ($s=1/3$).
Under such a tight bound of $x_4^{(0)}$, the contribution 
from $x_4$ to the total energy density at low redshifts is
 significantly suppressed relative to $|x_1|, x_2,x_3$. 
Hence, for the theories with $s \neq -1$, the quartic 
derivative interaction $a_4X^2$ is practically irrelevant to 
the dynamics of dark energy after the matter-dominated 
epoch. We caution, however, that $x_4$ tends to dominate 
over $|x_1|, x_2,x_3$ as we go back to the past 
(see Figs.~\ref{fig1} and \ref{fig2}), so it can affect 
the stability conditions in the deep radiation era 
even with the value of $x_4^{(0)}$ of order $10^{-14}$.

The quantity (\ref{Qs}) associated with the no-ghost condition of scalar perturbations yields 
\be
Q_s=\frac{3(5-x_4)q_s}
{200\pi G (x_3+2x_4-2)^2}\,,
\ee
where $q_s$ is defined by Eq.~(\ref{qs}).
Under the condition (\ref{tennoghost}), the 
scalar ghost is absent for 
\be
q_s>0\,,
\label{qscon}
\ee
which does not depend on $s$.
In Fig.~\ref{fig4}, we plot the evolution of the quantity $Q_s$ 
for the background cosmology corresponding to Fig.~\ref{fig1}. 
The quantity $Q_s$ remains positive throughout the cosmic 
expansion history, by reflecting the fact that the contributions 
to $Q_s$ from positive values of $x_2,x_3,x_4$ dominate 
over $|x_1|$.

For the scalar propagation speed squared (\ref{cs}), we first 
consider the regime in which the condition 
$x_4 \gg \{|x_1|, x_2, x_3 \}$ is satisfied in the 
early Universe. In this case, we obtain
\be
c_s^2 \simeq \frac{1}{36} \left[ 5+s
+(3-5s)\Omega_r \right]\,,
\label{cs4}
\ee
so that $c_s^2 \simeq (2-s)/9$ during the radiation dominance 
and $c_s^2 \simeq (5+s)/36$ during the matter dominance, respectively.  
To avoid the Laplacian instability in these 
epochs, we require that the ratio $s=b_4/a_4$ needs to be 
in the range
\be
-5 \le s \le 2\,.
\label{srange}
\ee
The model with $c_t^2=1$ corresponds to $s=-1$, 
so it satisfies the condition (\ref{srange}).  
In Fig.~\ref{fig4}, we plot the evolution of $c_s^2$ 
for $s=-1$ with the background initial conditions 
same as those in Fig.~\ref{fig1}.
We observe that $c_s^2$ starts to 
evolve from the value close to $1/3$ in the deep 
radiation era and decreases according to 
$c_s^2 \simeq (1+2\Omega_r)/9$ by the moment 
at which $x_3$ dominates over $x_4$.

In the regime characterized by $x_3 \gg \{|x_1|, x_2, x_4 \}$, 
Eq.~(\ref{cs}) reduces to
\be
c_s^2 \simeq \frac{1}{12} \left( 5+\Omega_r \right)\,,
\label{cs3}
\ee
which is positive.
In Fig.~\ref{fig4}, we can confirm that, during the matter dominance, 
$c_s^2$ temporally approaches the value $5/12$ 
after $x_3$ dominates over $x_4$.

On the de Sitter attractor, there are two relations given by 
Eq.~(\ref{dSre}) with $\Omega_r=0$. In this case, the 
quantity $\beta_{\rm H}$ in Eq.~(\ref{cs}) is equivalent 
to 0 and hence 
\be
c_s^2=c_{\rm H}^2
=-\frac{(6+6x_1+x_2)[18+9x_1+s(3x_1+x_2+3)
(6+15x_1+4x_2)]}
{3(3x_1+x_2)[4x_2^2+6x_2-36+3x_1(5x_2-6)]}\,,
\label{csds}
\ee
which is required to be positive to ensure the Laplacian stability 
on the de Sitter fixed point. 
In the numerical simulation of Fig.~\ref{fig4},
the asymptotic future values of $x_1$ and $x_2$ are 
given, respectively, 
$x_1=-1.96556397$ and $x_2=2.89669243$, so 
that $c_s^2=2.84 \times 10^{-3}>0$ from 
Eq.~(\ref{csds}). 
In Fig.~\ref{fig4}, we observe that $c_s^2$ is
positive from the radiation era to the de Sitter attractor.
The other propagation speed squared $\tilde{c}_r^2$ is 
always close to $1/3$. 
Thus, the model discussed above suffers neither 
ghost nor Laplacian instabilities.

We also computed the quantities $Q_t$, $Q_s$, $c_s^2$, 
and $\tilde{c}_r^2$ for all the four cases shown in 
Fig.~\ref{fig3} and confirmed that, for $s=-1$,  
they remain to be positive during the cosmic expansion 
history. Thus, in our model, it is possible to realize  
a wide variety of the dark energy equation of state 
including the entry to the  
region $w_{\rm DE}<-1$, while 
avoiding the appearance of ghosts and Laplacian instabilities.  
For larger initial conditions of $x_4$, there is a tendency that 
the scalar propagation speed squared temporally 
enters the region $c_s^2<0$ right after the end of 
the dominance of $x_4$ over $|x_1|, x_2, x_3$. 
To avoid this behavior, today's values $x_4$ 
are typically in the range 
\be
x_4^{(0)} \lesssim 10^{-4}\,.
\label{x4com}
\ee
The numerical simulation of Fig.~\ref{fig2} corresponds to 
the marginal case in which the condition $c_s^2>0$ 
is always satisfied. Since we have not searched for the whole parameter space of initial conditions, 
the condition (\ref{x4com}) should be regarded 
only as a criterion for avoiding $c_s^2<0$. 
To derive precise bounds on $x_4^{(0)}$, we need to 
carry out the likelihood analysis by considering 
all possible initial conditions.

\section{Growth of linear cosmological perturbations}
\label{vasec} 

We study the evolution of linear perturbations relevant to the observations of RSDs and weak lensing for 
the dark energy model given by the functions (\ref{model}). 
Since we are interested in the growth  
of matter perturbations long after the end of the radiation era, 
we will neglect the radiation and 
consider nonrelativistic matter alone satisfying 
$w_m=0$ and $c_m^2=0$.
The parameters $\alpha_{\rm M}$, 
$\alpha_{\rm B}$, $\alpha_{\rm K}$, and 
$\alpha_{\rm H}$ defined in Eq.~(\ref{aldef}) 
reduce, respectively, to 
\be
\alpha_{\rm M}=-\frac{4x_4 \epsilon_{\phi}}{5-x_4}\,,
\qquad 
\alpha_{\rm B}=-\frac{5x_3+8x_4}{2(5-x_4)}\,,
\qquad 
\alpha_{\rm K}=\frac{6(5x_1+10x_2+5x_3+6x_4)}
{5-x_4}\,,
\qquad
\alpha_{\rm H}=\frac{(1-3s)x_4}{5-x_4}\,.
\label{alM}
\ee
{}From Eqs.~(\ref{per1})-(\ref{per5}), the perturbations 
$\zeta$, $\chi \equiv H \psi$, $\delta$, and $V_m$ obey 
the first-order differential equations
\ba
\zeta' 
&=& \frac{5[3\Omega_m V_m+(2-x_3-2x_4) 
\delta N]}{2(5-x_4)}\,,
\label{zetaeq}\\
\chi'
&=& \frac{5(h-1)+(1-h+4\epsilon_{\phi})x_4}{5-x_4}
\chi -\left( 1+\alpha_{\rm H} \right)\delta N-c_t^2\zeta\,,\\
\delta'
&=& -\frac{15[3\Omega_m V_m+(2-x_3-2x_4) 
\delta N]}{2(5-x_4)}+{\cal K}^2 \left( V_m-\chi \right)\,,\\
V_m'
&=& hV_m-\delta N\,,
\label{Vmeq}
\ea
where ${\cal K} \equiv k/(aH)$, and 
\be
\delta N=
\frac{30\Omega_m(5-x_4)\delta-4(1+\alpha_{\rm H})
(5-x_4)^2{\cal K}^2 \zeta-5(2-x_3-2x_4)[2(5-x_4){\cal K}^2\chi
+45\Omega_mV_m]}
{60(5-x_4)(x_1+2x_2)+75x_3(4+x_3)+180(x_3+2)
x_4+120x_4^2}\,.
\label{delN}
\ee
Taking the ${\cal N}$ derivative of Eq.~(\ref{Vmeq}) and 
using other perturbation equations of motion, we obtain
\be
V_m''+\alpha_1V_m'+\alpha_2 V_m
=\beta_1 \chi+\beta_2 \zeta\,,
\label{Vmse}
\ee
where $\alpha_1, \alpha_2, \beta_1, \beta_2$ are 
time-dependent functions. 
The general solution to Eq.~(\ref{Vmse}) can be written 
as the sum of the homogenous solution to Eq.~(\ref{Vmse}) 
and the special solution induced by the r.h.s. of Eq.~(\ref{Vmse}). 
The deviation from Horndeski theories generally gives 
rise to a term proportional to $\alpha_{\rm H}\Omega_m{\cal K}^2$ 
in $\alpha_2$ \cite{Koyama}. 
In the present model, such a term (denoted as 
$\tilde{\alpha}_2$) is given by 
\be
\tilde{\alpha}_2=-\frac{2}{q_s} 
\left(5-x_4 \right) 
\alpha_{\rm H}\Omega_m {\cal K}^2\,.
\label{tal2}
\ee
Under the conditions (\ref{tennoghost}) and (\ref{qscon}), 
$\tilde{\alpha}_2$ is negative for 
$\alpha_{\rm H}>0$, whereas 
$\tilde{\alpha}_2>0$ for $\alpha_{\rm H}<0$. 
If $x_4$ dominates over $|x_1|,x_2,x_3$, then the quantity 
(\ref{tal2}) approximately reduces to 
$\tilde{\alpha}_2 \simeq -(1-3s)\Omega_m {\cal K}^2/12$, 
so the coefficient in front of the ${\cal K}^2$ term is 
smaller than the order of unity for $|s| \lesssim {\cal O}(1)$.
If either of $|x_1|, x_2, x_3$ dominates over $x_4$, the 
coefficient is much smaller than the order of 1. 
Thus, unlike the model of Ref.~\cite{Koyama}, the 
perturbation $V_m$ is not plagued by the problem 
of a heavy oscillation arising from a large positive 
coefficient in front of $\alpha_{\rm H}\Omega_m{\cal K}^2$. 

If $\tilde{\alpha}_2$ is negative, one may worry that the term 
$\tilde{\alpha}_2$ can lead to the instability in the small-scale 
limit (${\cal K} \to 0$).  
However, the coefficients $\beta_1$ and $\beta_2$ 
also contain contributions proportional 
to ${\cal K}^2$, so the term $\tilde{\alpha}_2 V_m$ 
can balance with the r.h.s. of Eq.~(\ref{Vmse}). 
This is consistent with the fact that the matter sound 
speed squared $\tilde{c}_m^2$ in the small-scale limit  
vanishes for $c_m^2 \to 0$, which implies the absence of 
Laplacian instabilities for matter perturbations.

In the following, we choose the initial conditions satisfying 
$\alpha_2V_m= \beta_1 \chi+\beta_2 \zeta$, 
$\zeta'=0$, and $\chi'=0$ in the deep matter-dominated 
epoch. {}From Eq.~(\ref{zetaeq}), the condition 
$\zeta'=0$ 
translates to $\delta N \simeq -3\Omega_m V_m/2$ for 
$x_3,x_4 \ll 1$, which corresponds to the same 
initial condition as that in GR. 
In this case, we obtain $V_m' \simeq 0$ from Eq.~(\ref{Vmeq}) in the deep matter era ($\Omega_m \simeq 1$ and $h \simeq -3/2$).
Since the initial condition 
$\alpha_2V_m=\beta_1 \chi+\beta_2 \zeta$ is chosen, 
we have $V_m'' \simeq 0$ from Eq.~(\ref{Vmse}). 
Using also the initial condition $\chi' \simeq 0$ mimicking the behavior 
of GR, the initial values of $V_m$, $\zeta$, and $\chi$ can be expressed 
in terms of $\delta$. The initial value of $\delta$ is chosen such that 
today's amplitude of $\delta_m$ is consistent with the bound 
constrained from CMB.

\begin{figure}[h]
\begin{center}
\includegraphics[height=3.3in,width=3.4in]{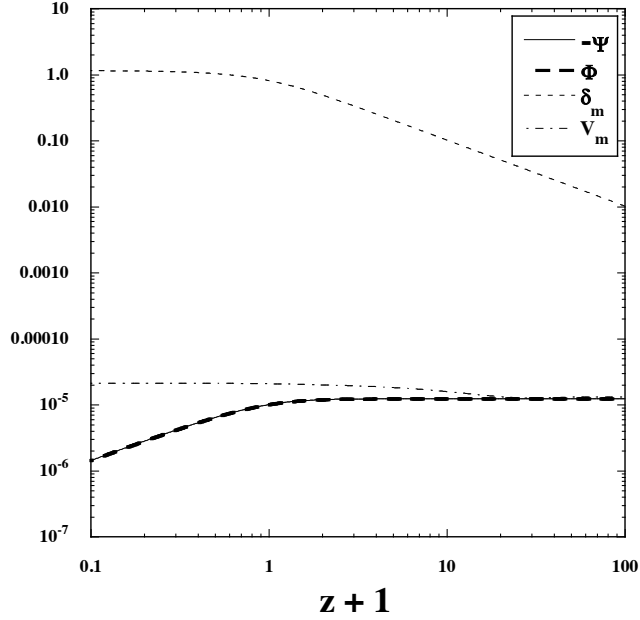}
\end{center}
\caption{\label{fig5}
Evolution of the two gravitational potentials 
$-\Psi$ and $\Phi$, the density contrast $\delta_m$, and 
the velocity potential $V_m$ versus $z+1$ for the model $c_t^2=1$.
The background cosmology corresponds to that in 
Fig.~\ref{fig1} without radiation.
We start to integrate the perturbation equations of motion 
(\ref{zetaeq})-(\ref{Vmeq}) with the initial conditions 
explained in the main text at the redshift $z=100$. 
Today's values of $\delta_m$ and ${\cal K}$ are
$\delta_m=0.815$ and ${\cal K}=200$, respectively.}
\end{figure}

In Fig.~\ref{fig5}, we plot the evolution of 
$-\Psi,\Phi,\delta_m,V_m$ for $s=-1$ (i.e., $c_t^2=1$) 
corresponding to the background cosmology of Fig.~\ref{fig1}. 
The initial conditions of perturbations are chosen as 
those explained above around the redshift $z=100$, 
with today's values $\delta_m(z=0)=0.815$ 
and ${\cal K}(z=0)=200$.
In the deep matter era, the velocity potential $V_m$ 
stays nearly constant without heavy oscillations, so that 
$V_m' \simeq 0$ and $V_m'' \simeq 0$ in Eq.~(\ref{Vmse}). 
The variation of $V_m$ starts to occur after the dark energy density 
contributes to the background density at low 
redshifts. Since the condition $x_4 \ll \{ |x_1|, x_2, x_3 \}$ is 
satisfied for $z \lesssim 100$, $\tilde{\alpha}_2$ is much smaller 
than the order of ${\cal K}^2$.

The numerical simulation of Fig.~\ref{fig5} shows that the two 
gauge-invariant gravitational potentials $-\Psi$ and 
$\Phi$ are almost identical to each other. 
This can be understood as follows. 
In the definition of $\alpha_{\rm M}$ in Eq.~(\ref{alM}), 
the quantity $\epsilon_{\phi}$ is at most of the order 1, 
so that $|\alpha_{\rm M}| \lesssim x_4 \ll 1$. 
As we see in Figs.~\ref{fig1} and \ref{fig2}, 
$x_3$ is much larger than $x_4$ at low redshifts 
and hence the quantity $|\alpha_{\rm B}|$ is of the order of $x_3$. 
Then, the ratio $|\alpha_{\rm M}/\alpha_{\rm B}|$ 
is in the range 
\be
\left| \frac{\alpha_{\rm M}}{\alpha_{\rm B}} \right| 
\lesssim 
\left| \frac{x_4}{x_3} \right| \ll 1\,. 
\ee
In the present model, the parameter $\alpha_{\rm H}$ 
appearing on the r.h.s. of Eqs.~(\ref{quasi1}) and 
(\ref{quasi3}) is much smaller than 1 at low redshifts, 
so the results (\ref{Gefff})-(\ref{Sigmaf}) derived under the 
quasi-static approximation should not lose their validity. 
Since we are now considering the case $c_t^2=1$, the 
gravitational slip parameter is given by Eq.~(\ref{eta2}).
Taking the limit $|\alpha_{\rm M}| \ll |\alpha_{\rm B}|$ 
in Eq.~(\ref{eta2}), we obtain
\be
\eta \simeq 1\,,
\label{eta1}
\ee
and hence $-\Psi \simeq \Phi$.
We note that taking the same limit in Eq.~(\ref{etaf}) 
leads to $\eta$ generally different from 1. 
This means that the condition $c_t^2=1$ plays an 
important role for realizing the value (\ref{eta1}).

\begin{figure}[h]
\begin{center}
\includegraphics[height=3.1in,width=3.3in]{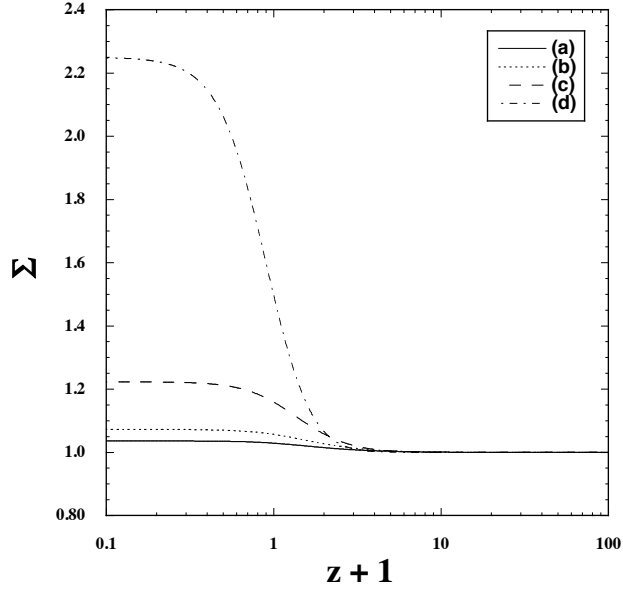}
\end{center}
\caption{\label{fig6}
Evolution of the quantity $\Sigma$ versus $z+1$ for the 
model $c_t^2=1$. Each curve corresponds to the background 
cosmologies shown as the cases (a), (b), (c), (d) 
in Fig.~\ref{fig3} without radiation. 
We start to integrate the perturbation 
equations of motion at the redshift $z=100$ 
by using the initial conditions of $\zeta,\chi,V_m$ 
satisfying $\zeta'=0$, $\chi'=0$, and
$V_m=(\beta_1 \chi+\beta_2 \zeta)/\alpha_2$.
Today's values of $\delta_m$ and ${\cal K}$ are 
$\delta_m=0.815$ and ${\cal K}=200$, respectively.}
\end{figure}

\begin{figure}[h]
\begin{center}
\includegraphics[height=3.1in,width=3.3in]{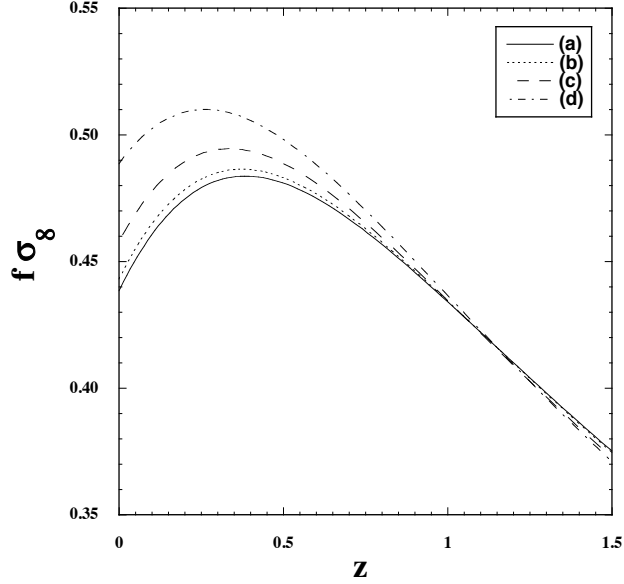}
\end{center}
\caption{\label{fig7}
Evolution of $f\sigma_8$ versus $z$ for the model 
$c_t^2=1$. Each curve corresponds to the cases 
(a), (b), (c), (d) explained in the caption of Fig.~\ref{fig6}.}
\end{figure}

On using the property (\ref{eta1}), it follows that the two 
quantities $\mu$ and $\Sigma$ are almost equivalent to each other. 
Indeed, taking the limit $|\alpha_{\rm M}| \ll |\alpha_{\rm B}|$ 
in Eqs.~(\ref{Gefff2}) and (\ref{Sigmaf2}) gives
\be
\mu \simeq \Sigma \simeq 
\frac{1}{16\pi G Q_t} \left[ 1+
\frac{2Q_t \alpha_{\rm B}^2}{Q_sc_s^2 (1+\alpha_{\rm B})^2} \right]\,.
\label{muSig}
\ee
Since $1/(16\pi GQ_t)=5/(5-x_4)>1$ for $x_4>0$, 
both $\mu$ and $\Sigma$ are larger than 1 under the 
conditions $Q_s>0$, $Q_t>0$, $c_s^2>0$. 
In Ref.~\cite{Pogosian}, it was shown that the observational 
data generally favor the region $(\mu-1)(\Sigma-1) \geq 0$, so 
our model is consistent with this property.
The parameter $\alpha_{\rm B}$, which can be estimated 
as $\alpha_{\rm B} \simeq -x_3/2$ for $x_3 \gg x_4$, 
gives rise to important modifications to 
$\mu$ and $\Sigma$ compared to those in GR.

In Fig.~\ref{fig6}, we show the evolution of $\Sigma$ 
for the four different backgrounds corresponding to those 
in Fig.~\ref{fig3}. 
We solved the full perturbation equations of motion from 
the redshift $z \simeq 100$ without 
employing the quasi-static approximation and 
found that the numerically derived values of 
$\mu$ and $\Sigma$ are very close to each other. 
Moreover, we confirmed that 
Eqs.~(\ref{Gefff2})-(\ref{Sigmaf2})
derived under the quasi-static approximation without 
the terms on the r.h.s. of Eqs.~(\ref{quasi1}) and 
(\ref{quasi3}) provide good estimates for the 
evolution of $\mu$ and $\Sigma$ by today.
In Fig.~\ref{fig6}, we observe that the deviation of $\Sigma$ 
from $1$ is more significant for the cases in which the dark energy equation of state largely deviates from $-1$ at lower redshifts (see Fig.~\ref{fig3}). 
This reflects the fact that today's value of  
$x_3$ (denoted as $x_3^{(0)}$)
tends to be larger for $w_{\rm DE}$ entering the 
region $w_{\rm DE}<-1$ at later epochs, e.g., 
$x_3^{(0)}=5.6 \times 10^{-2}$ and $x_3^{(0)}=0.82$ 
for the cases (a) and (d), respectively. 
For larger $x_3^{(0)}$, the quantities $\mu$ and $\Sigma$ 
increase according to Eq.~(\ref{muSig}). 

In Fig.~\ref{fig7}, we plot the evolution of $f\sigma_8$ 
corresponding to the cases (a)-(d) in Fig.~\ref{fig6},  
where $f \equiv \dot{\delta}_m/(H\delta_m)$ and 
$\sigma_8$ is the amplitude of over-density at the 
comoving $8\,h^{-1}$ Mpc scale 
($h$ is today's normalized Hubble parameter 
$H_0 = 100\,h$~km sec$^{-1}$\,Mpc$^{-1}$).
Today's value of $\sigma_8$ is chosen to be
$\sigma_8(z=0)=0.815$ for the consistency with the 
constraint from CMB \cite{Planckdark,Planck15}.
Since the normalized effective gravitational coupling 
$\mu=G_{\rm eff}/G$ at low redshifts increases for larger 
$x_3^{(0)}$, it is possible to distinguish the several different 
cases plotted in Fig.~\ref{fig7} from the RSD 
measurements (see, e.g., Ref.~\cite{Okumura}).
We note that the cubic coupling $G_3=3a_3X$ is crucially 
important to modify the values of $\mu$ and $\Sigma$ 
compared to those in GR. 
It remains to be seen how the observation data of RSDs and 
weak lensing place constraints on the values $x_3^{(0)}$ 
and $x_4^{(0)}$.

\section{Constraints on $\alpha_{\rm H}$ from 
massive objects}
\label{astrosec} 

Finally, we discuss constraints on our dark energy model
arising from the change of gravitational law inside local 
massive objects. In such regimes, the nonlinear 
beyond-Horndeski derivative interaction leads to modifications to the gravitational potentials of GR \cite{Koba15,Sakstein,Mizuno,Babichev16} in a way different from those derived for linear cosmological perturbations 
in Sec.~\ref{vasec}.
Hence it is possible to place constraints on the parameter 
$\alpha_{\rm H}$ from several astrophysical observations 
associated with nonrelativistic compact objects.

Let us first consider a scalar field $\phi$ with the radial dependence $r$ alone around a spherically symmetric and static body. For the model in which the constant difference 
between $-A_4$ and $B_4$ is present with
the beyond-Horndeski interaction $F_4 \propto X^{-2}$ 
\cite{Koyama}, the scalar curvature exhibits a divergence 
of the form $R=-2\alpha_{\rm H}/r^2$ at 
$r=0$ \cite{DKT15,DKT15b}.
This is attributed to the appearance of a solid angle deficit singularity, which is related to the violation of 
the geometric structure of space-time. 
In our model (\ref{G234}), which corresponds to 
$F_4={\rm constant}$, the parameter $\alpha_{\rm H}$ 
is proportional to $X^2$. 
Since the regularity of $r$-dependent scalar field
requires that $d\phi/dr=0$ at $r=0$, this means that 
the parameter $\alpha_{\rm H}$, which is proportional 
to $(d\phi/dr)^4$, vanishes at the center of body.
Hence our model is free from the problem of the solid 
angle deficit singularity (as in the model studied 
in Ref.~\cite{KTF15}).

More importantly, in our model, the field kinetic energy $X$ 
drives the cosmic acceleration.
In this case, $X$ contains the cosmological time derivative 
$(d\bar{\phi}/dt)^2$ as well as the radial derivative 
$(d\chi/dr)^2$ around a local source.
This situation is analogous to what happens 
in generalized Proca theories, in which 
a temporal vector component $\phi^2$ is present  
besides a longitudinal scalar mode 
$(d\chi/dr)^2$ \cite{screeningProca}.
In such cases, it was shown that the solid angle deficit singularity is generally absent by the dominance 
of $\phi^2$ over $(d\chi/dr)^2$ in the limit 
that $(d\chi/dr)^2 \to 0$ \cite{HKT16}. 
In GLPV theories where the time derivative of the scalar 
leads to cosmic acceleration, the problem of the 
solid angle deficit singularity
does not arise either.

In GLPV theories, the corrections to gravitational potentials 
$\Psi$ and $\Phi$ inside massive objects induced by the beyond-Horndeski nonlinear 
derivative interaction were already derived in Refs.~\cite{Koba15,Sakstein,Mizuno,Babichev16,Langlois17,Dima17}. We consider the Newtonian gravitational potentials 
$\Psi(r)$ and $\Phi(r)$ with the line 
element\footnote{For the consistency with 
Eq.~(\ref{Psidef}), we have defined $\Psi(r)$ 
and $\Phi(r)$ different from those used in 
the literature. Compared to Refs.~\cite{Koba15,Sakstein,Mizuno,Babichev16,Langlois17,Dima17}, there is the correspondence: 
$\Psi(r) \to \Phi(r)$ and $\Phi(r) \to -\Psi(r)$.}
\be
ds^2=-\left[ 1+2\Psi(r) \right] dt^2+ 
\left[ 1+2\Phi (r) \right] \delta_{ij}dx^idx^j\,,
\ee
where $r=\sqrt{\delta_{ij}x^ix^j}$ is the radial 
coordinate. 
The scalar field $\phi$ and the matter density 
$\rho_m$ acquire perturbations by the existence of a compact object, as $\phi=\bar{\phi}(t)+\chi(r)$ and 
$\rho_m=\bar{\rho}_m(t)+\delta \rho_m(r)$. 
The time derivative $\dot{\bar{\phi}}(t)$, which is 
the source for dark energy, works as 
a cosmological boundary term.

On small scales relevant to the physics of compact 
objects, the spatial derivatives of perturbations dominate 
over their time derivatives. We neglect such time 
derivatives and keep all the terms with second-order 
derivatives preserving the Galileon symmetry.
Using the equation of motion for $\chi$, 
the field derivative $d\chi/dr$ in the region of over density 
can be expressed in terms of the mass function 
${\cal M}(r)=\int_0^{r} 4\pi \tilde{r}^2 \delta \rho_m 
(\tilde{r}) d\tilde{r}$ and its $r$ derivative \cite{Koba15}. 
Substituting this relation into the equations of motion 
for $\Psi$ and $\Phi$, their $r$-derivatives can be 
written as \cite{Dima17}
\ba
\frac{d\Psi(r)}{dr} &=& G_{\rm N} \left[ \frac{{\cal M}(r)}{r^2}
+\gamma_1 \frac{d^2{\cal M}(r)}{dr^2} \right]\,,
\label{lo1}\\
\frac{d\Phi(r)}{dr} &=& -G_{\rm N} \left[ \frac{{\cal M}(r)}{r^2}
+\gamma_2 \frac{1}{r}\frac{d{\cal M}(r)}{dr} \right]\,,
\label{lo2}
\ea
where 
\ba
G_{\rm N} &=&\frac{1}{16\pi (-A_4)(1+\alpha_{\rm V})}\,,\label{GN} \\
\gamma_1 &=& \frac{\alpha_{\rm H}^2}
{c_t^2(1+\alpha_{\rm V})-1-\alpha_{\rm H}}\,,\label{gamma1}\\
\gamma_2 &=&  \frac{\alpha_{\rm H}(\alpha_{\rm V}-\alpha_{\rm H})}
{c_t^2(1+\alpha_{\rm V})-1-\alpha_{\rm H}}\,,\label{gamma2}
\ea
with 
\be
\alpha_{\rm V}=-\frac{A_{4,N}}{A_4}\,.
\label{alV}
\ee
In Refs.~\cite{Sakstein,Mizuno,Babichev16}, the authors studied the case of quartic GLPV theories without the cubic Lagrangian ($x_3=0$), 
so the quantity $\alpha_{\rm B}$ in Eq.~(\ref{alM}) is equivalent to $\alpha_{\rm V}=-4x_4/(5-x_4)$. 
Now, we are considering more general theories 
with $x_3 \neq 0$, in which case the terms $\alpha_{\rm V}$ 
appearing in Eqs.~(\ref{GN})-(\ref{gamma2}) 
cannot be replaced with $\alpha_{\rm B}$.
Since $-A_4=(5-x_4)/(80\pi G)$ in our model, 
the effective Newton constant is given by 
\be
G_{\rm N}=\frac{G}{1-x_4}\,.
\ee
This is different from the effective gravitational 
coupling $G_{\rm eff}$ derived for {\it linear} 
matter density perturbations, see Eq.~(\ref{muSig}).
The gravitational coupling $G_{\rm eff}$, which contains 
$\alpha_{\rm B}$, is affected by the cubic 
derivative coupling $G_3=3a_3X$, while 
this is not the case for $G_{\rm N}$. 
The latter property is attributed to the fact that, 
apart from the beyond-Horndeski interaction,  
the propagation of fifth forces is suppressed by
 the operation of the 
Vainshtein mechanism.

For the model with $s=-1$, we have
\be
c_t^2=1\,,\qquad
\alpha_{\rm H}=-\alpha_{\rm V}=\frac{4x_4}{5-x_4}\,,
\ee
where we used Eq.~(\ref{alM}).
Hence the two parameters (\ref{gamma1}) 
and (\ref{gamma2}) reduce, respectively, to 
\be
-2\gamma_1=\gamma_2=\alpha_{\rm H}
=\frac{4x_4}{5-x_4}\,.
\ee

In GR, the two gravitational potentials $-\Psi$ and $\Phi$ are 
equivalent to each other. This equivalence is broken in GLPV 
theories due to the existence of 
two terms containing $\gamma_1$ and 
$\gamma_2$ in Eqs.~(\ref{gamma1})-(\ref{gamma2}). 
The mass of galaxy clusters measured by weak lensing (sensitive 
to $\psi_{\rm eff}=\Phi-\Psi$) and X-ray observations (sensitive to 
$-\Psi$) does not coincide for $\alpha_{\rm H} \neq 0$, so it is possible 
to place bounds on the values of $\gamma_1$ and $\gamma_2$.
{}From the CFHTLenS weak lensing data and the X-ray data from 
XMM-Newton in the redshift range $0.1<z<1.2$, 
these parameters are constrained to be 
$-0.78 \le 4\gamma_1 \le 0.82$ and 
$-1.41 \le -4\gamma_2/5 \le 1.00$ \cite{Sakstein3}. 
These constraints translate to 
\be
-0.57 \le x_4 \le 0.44\,,\quad {\rm and} \quad 
-2.27 \le x_4 \le 1.53\,,
\label{bound1}
\ee
respectively. 

There are also bounds on $\gamma_1$ arising from the modification to  
the mass of nonrelativistic stars. For the consistency of the white dwarf 
of lowest mass with the Chandrasekhar limit and  the consistency of the 
minimum mass for hydrogen burning in stars with the red dwarf of
lowest mass\footnote{We updated 
the upper bound of $4\gamma_1$ according to the arXiv version 3 
of Ref.~\cite{Sakstein1}. 
We thank Jeremy Sakstein for pointing it out.}, the parameter $\gamma_1$ is constrained to be 
$-0.51<4\gamma_1<1.6$ \cite{Sakstein1,Sakstein2}. 
Then, we obtain the bound

\be
-1.25 \le x_4 \le 0.30\,,
\label{bound2}
\ee
at $z=0$.

There is also a constraint on $x_4$ from the binary pulsars.  
The orbital period of binary stars is proportional to $(-A_4G_{\rm N}c_t)^{-1}$, so for the model with $c_t^2=1$, 
it is simply related to the parameter 
$\gamma_0 \equiv 1+\alpha_{\rm V}$. 
The observations of Hulse-Taylor binary pulsar PSR B1913+16 
provided the bound $-7.5 \times 10^{-3} \le \gamma_0-1 
\le 2.5 \times 10^{-3}$ \cite{Hulse,Dima17}, 
so this translates to 
\be
-0.0031 \le x_4 \le 0.0094\,,
\label{bound3}
\ee
at $z \ll 1$. 

The constraints (\ref{bound1}), (\ref{bound2}), and (\ref{bound3}) need to be satisfied in the redshift 
range $z \lesssim 1$. 
The tightest one comes from binary pulsars, so that all the astrophysical bounds
discussed above can be satisfied for $|x_4| \lesssim 10^{-3}$ at $z \lesssim 1$. 
This is weaker than the criterion (\ref{x4com}) derived for the realization 
of cosmological dynamics without theoretical inconsistencies.

For the models with $s \neq -1$, the quantities $\gamma_0$, $\gamma_1$, 
and $\gamma_2$ are given, respectively, by 
\be
\gamma_0-1=-\frac{4x_4}{5-x_4}\,,\qquad
\gamma_1=-\frac{x_4(1-3s)^2}{20(1-s)+8x_4s}\,,\qquad 
\gamma_2=\frac{x_4(5-3s)(1-3s)}{20(1-s)+8x_4s}\,.
\ee
Since we are now considering the case $c_t^2 \neq 1$, 
today's value of $x_4$ is constrained to be smaller than the order of 
$10^{-14}$ from the speed of gravitational waves. 
Then, the binary pulsar bound on $\gamma_0$ is trivially satisfied.
Provided that $s$ is not very close to 1, both $|\gamma_1|$ 
and $|\gamma_2|$ are at most of the order of $x_4$, 
so the other constraints on $\gamma_1$ and $\gamma_2$ 
are also satisfied.

\section{Conclusions}
\label{consec} 

In this paper, we proposed a dark energy model in the framework of GLPV theories consistent with the recent GW170817 bound of the tensor propagation 
speed. The existence of a beyond-Horndeski quartic 
coupling $F_4$ allows the possibility for realizing 
$c_t^2=1$ on the isotropic cosmological background 
under the condition (\ref{ctcon}).
Our model contains the Galileon interactions up to 
quartic order as well as the $a_2X^2$ term in the quadratic 
Lagrangian $G_2$. In the 3+1 ADM language, 
the model is given by the Lagrangian (\ref{Lag}) 
with the functions (\ref{model}). 
For the model with $s=b_4/a_4=-1$,
the tensor propagation speed squared in 
Eq.~(\ref{aH}) is equivalent to 1. 

In Sec.~\ref{backpersec}, we presented general background 
and scalar perturbation equations of motion in quartic-order GLPV theories without specifying the model.
We also revisited conditions for the absence of ghosts and 
Laplacian instabilities and the behavior of perturbations 
for the modes deep inside the (sound) horizon.
The deviation from Horndeski theories, weighed by 
the parameter $\alpha_{\rm H}$, generates
additional time derivatives to the scalar perturbation equations 
even under the quasi-static approximation. 
In the case where $|\alpha_{\rm H}|$ is much smaller than 1, 
we derived general expressions of two quantities 
$\mu$ and $\Sigma$ by using the dimensionless variables given by Eq.~(\ref{aldef}). 
If $c_t^2=1$, $\mu$ and $\Sigma$ 
reduce, respectively, to (\ref{Gefff2}) and (\ref{Sigmaf2}), 
so they are different from the value $1/(16\pi G Q_t)$ for 
the theories with $\alpha_{\rm B} \neq \alpha_{\rm M}$. 
The necessary condition for realizing 
$\mu=G_{\rm eff}/G<1$ corresponds to $Q_t>1/(16\pi G)$.

In Sec.~\ref{cosmosec}, we studied the background cosmological dynamics for the dark energy model 
given by the functions (\ref{model}). 
There exists a de Sitter fixed point satisfying the conditions 
(\ref{dSre}), which is always a stable attractor. 
In the early Universe, the cubic and quartic Galileon 
couplings with the density parameters $x_3$ and $x_4$ 
provide important contributions to 
dark energy relative to the k-essence density parameters 
$x_1$ and $x_2$.
In the radiation and matter eras, the dark energy equation 
of state is given by $w_{\rm DE} \simeq -\Omega_r/9$ 
for $x_4 \gg \{|x_1|, x_2, x_3 \}$, whereas 
$w_{\rm DE} \simeq 1/4-\Omega_r/12$ for 
$x_3 \gg \{|x_1|, x_2, x_4\}$. 
Depending on the moment of the dominance of $x_3$ 
over $x_4$ and on the values of $|x_1|$ and $x_2$, 
the dark energy equation of state either 
enters the regime $w_{\rm DE}<-1$ or stays in the 
region $w_{\rm DE}>-1$ (see Figs.~\ref{fig1} and \ref{fig2}).
For smaller $x_2$, the dominance of the $a_1X^2$ term in 
$G_2$ tends to occur at later epochs, so the larger deviation 
of $w_{\rm DE}$ from $-1$ occurs at lower redshifts 
(see Fig.~\ref{fig3}).

In Sec.~\ref{cosmosec}, we also showed that the tensor 
ghost and Laplacian instability are absent 
under the conditions (\ref{tennoghost}) and (\ref{tennoghost2}). {}From the observational bound of 
$c_t$, today's value of $x_4$ is constrained to be
$|(s+1)x_4^{(0)}| \lesssim 10^{-14}$. 
To avoid the Laplacian instability 
of scalar perturbations in the radiation and matter eras, 
the parameter $s$ needs to be in the range 
$-5 \le s \le 2$, which includes the case $s=-1$.
For the background cosmologies plotted 
in Fig.~\ref{fig3}, the model with $s=-1$
is plagued by neither ghost nor Laplacian instabilities 
throughout the cosmic expansion history.
With increasing $x_4^{(0)}$, 
there is a tendency that the scalar propagation speed 
squared temporally enters the region $c_s^2<0$ 
after the end of dominance of $x_4$ over 
$|x_1|, x_2, x_3$. To avoid this behavior, today's value 
of $x_4$ is typically in the range 
$x_4^{(0)} \lesssim 10^{-4}$.

In Sec.~\ref{vasec}, we studied the evolution of linear 
cosmological perturbations for our dark energy model with $s=-1$ and showed that the velocity potential $V_m$ does not 
suffer from the problem of a heavy oscillation induced by the 
beyond-Horndeski term $\alpha_{\rm H}$.
Since $|\alpha_{\rm H}| \ll 1$ in the matter era, 
the analytic estimations (\ref{Gefff2})-(\ref{Sigmaf2}) 
provide good approximations for discussing the 
evolution of gravitational potentials and matter 
density perturbations.
Since the condition $|\alpha_{\rm M}| \ll |\alpha_{\rm B}|$ 
is satisfied in our model, the gravitational slip parameter 
$\eta$ is close to 1 and hence $\mu \simeq \Sigma > 1$ 
for $x_4>0$. The deviations of $\mu$ and $\Sigma$ 
from those in GR ($\mu=\Sigma=1$) mostly come from the density parameter $x_3$ of cubic Galileons 
in $\alpha_{\rm B}$. As we see in Figs.~\ref{fig6} 
and \ref{fig7}, the cosmic growth rate tends to be larger 
for the cases in which the deviation of $w_{\rm DE}$
from $-1$ occurs significantly at lower redshifts.

In Sec.~\ref{astrosec}, we discussed constrains on 
$\alpha_{\rm H}$ arising from the change of 
gravitational law inside massive objects. 
The beyond-Horndeski nonlinear derivative interaction 
gives rise to modifications to the radial derivatives of  gravitational potentials 
$\Psi$ and $\Phi$ in the forms (\ref{lo1})-(\ref{lo2}).
For the model with $c_t^2=1$, the tightest bound on $x_4$ 
in the redshift range $0 \le z \lesssim 1$ comes from the 
orbital period of Hulse-Taylor binary 
pulsar, such that $-0.0031 \le x_4 \le 0.0094$. 
This bound is consistent with the criterion 
$x_4^{(0)} \lesssim 10^{-4}$ derived for 
the realization of viable cosmological solutions 
without Laplacian instabilities.

We have thus shown that our proposed model in GLPV theories offers interesting possibilities 
for realizing a wide variety of the dark energy equation of state without theoretical pathologies, 
while satisfying the observational bound of $c_t$. 
Moreover, the model gives rise to the distinguished cosmic 
growth history characterized by 
$\mu \simeq \Sigma>1$, which can be tested with 
future high-precision observational data of 
RSDs and weak lensing. 
It will be of interest to put precise observational
bounds on today's density parameters $x_3$ and $x_4$ 
to understand how much extent the Galileon and beyond-Horndeski interactions can contribute to the physics 
of late-time 
cosmic acceleration.

\section*{ACKNOWLEDGEMENTS}
RK is supported by the Grant-in-Aid for Young Scientists B 
of the JSPS No.\,17K14297. 
ST is supported by the Grant-in-Aid for Scientific Research 
Fund of the JSPS No.~16K05359 and 
MEXT KAKENHI Grant-in-Aid for 
Scientific Research on Innovative Areas ``Cosmic Acceleration'' (No.\,15H05890).



\begin{thebibliography}{10}

\bibitem{SN1}
A.~G.~Riess \textit{et al.},
Astron.\ J.\  {\bf 116}, 1009 (1998) 
[astro-ph/9805201].

\bibitem{SN2}
S.~Perlmutter \textit{et al.},
Astrophys.\ J.\  {\bf 517}, 565 (1999) 
[astro-ph/9812133].

\bibitem{WMAP}
D.~N.~Spergel {\it et al.},
Astrophys.\ J.\ Suppl.\  {\bf 148}, 175 (2003)
[astro-ph/0302209].

\bibitem{Planck}
P.~A.~R.~Ade {\it et al.},
Astron.\ Astrophys.\  {\bf 571}, A16 (2014)
[arXiv:1303.5076 [astro-ph.CO]].

\bibitem{BAO}
D.~J.~Eisenstein {\it et al.},
Astrophys.\ J.\  {\bf 633}, 560 (2005)
[astro-ph/0501171].

\bibitem{Planckdark} 
P.~A.~R.~Ade {\it et al.} [Planck Collaboration],
Astron.\ Astrophys.\  {\bf 594}, A14 (2016)
[arXiv:1502.01590 [astro-ph.CO]].

\bibitem{review1}
E.~J.~Copeland, M.~Sami and S.~Tsujikawa,
Int.\ J.\ Mod.\ Phys.\ D {\bf 15}, 1753 (2006)
[hep-th/0603057].

\bibitem{review2}
A.~Silvestri and M.~Trodden,
Rept.\ Prog.\ Phys.\  {\bf 72}, 096901 (2009)
[arXiv:0904.0024 [astro-ph.CO]]

\bibitem{review3}
T.~P.~Sotiriou and V.~Faraoni,
Rev.\ Mod.\ Phys.\  {\bf 82}, 451 (2010)
[arXiv:0805.1726 [gr-qc]].

\bibitem{review4}
A.~De Felice and S.~Tsujikawa,
Living Rev.\ Rel.\  {\bf 13}, 3 (2010)
[arXiv:1002.4928 [gr-qc]].

\bibitem{review5}
T.~Clifton, P.~G.~Ferreira, A.~Padilla and C.~Skordis,
Phys.\ Rept.\  {\bf 513}, 1 (2012)
[arXiv:1106.2476 [astro-ph.CO]].
  
\bibitem{review6} 
A.~Joyce, B.~Jain, J.~Khoury and M.~Trodden,
Phys.\ Rept.\  {\bf 568}, 1 (2015)
[arXiv:1407.0059 [astro-ph.CO]].

\bibitem{quin1}
Y.~Fujii, Phys.\ Rev.\ D {\bf 26}, 2580 (1982).

\bibitem{quin2}
L.~H.~Ford,
Phys.\ Rev.\ D {\bf 35}, 2339 (1987).

\bibitem{quin3}
C.~Wetterich, Nucl. \ Phys \ B. {\bf 302}, 668 (1988).

\bibitem{quin4}
T.~Chiba, N.~Sugiyama and T.~Nakamura,
Mon.\ Not.\ Roy.\ Astron.\ Soc.\  {\bf 289}, L5 (1997)
[astro-ph/9704199].

\bibitem{quin5}
P.~G.~Ferreira and M.~Joyce,
Phys.\ Rev.\ Lett.\  {\bf 79}, 4740 (1997)
[astro-ph/9707286].

\bibitem{quin6}
R.~R.~Caldwell, R.~Dave and P.~J.~Steinhardt,
Phys.\ Rev.\ Lett.\  {\bf 80}, 1582 (1998).
[astro-ph/9708069].

\bibitem{kes1} 
C.~Armendariz-Picon, T.~Damour and V.~F.~Mukhanov,
Phys.\ Lett.\ B {\bf 458}, 209 (1999)
[hep-th/9904075].

\bibitem{kes2} 
T.~Chiba, T.~Okabe and M.~Yamaguchi,
Phys.\ Rev.\ D {\bf 62}, 023511 (2000)
[astro-ph/9912463].

\bibitem{kes3} 
C.~Armendariz-Picon, V.~F.~Mukhanov and P.~J.~Steinhardt,
Phys.\ Rev.\ Lett.\  {\bf 85}, 4438 (2000)
[astro-ph/0004134].

\bibitem{CDT} 
T.~Chiba, A.~De Felice and S.~Tsujikawa,
Phys.\ Rev.\ D {\bf 87}, no. 8, 083505 (2013)
[arXiv:1210.3859 [astro-ph.CO]].

\bibitem{DHT17} 
A.~De Felice, L.~Heisenberg and S.~Tsujikawa,
Phys.\ Rev.\ D {\bf 95}, 123540 (2017)
[arXiv:1703.09573 [astro-ph.CO]].

\bibitem{Horndeski} 
G.~W.~Horndeski,
Int.\ J.\ Theor.\ Phys.\  {\bf 10}, 363 (1974).

\bibitem{Horn1} 
C.~Deffayet, X.~Gao, D.~A.~Steer and G.~Zahariade,
Phys.\ Rev.\ D {\bf 84}, 064039 (2011)
[arXiv:1103.3260 [hep-th]].
  
\bibitem{Horn2} 
T.~Kobayashi, M.~Yamaguchi and J.~'i.~Yokoyama,
Prog.\ Theor.\ Phys.\  {\bf 126}, 511 (2011)
[arXiv:1105.5723 [hep-th]].

\bibitem{Horn3} 
C.~Charmousis, E.~J.~Copeland, A.~Padilla and P.~M.~Saffin,
Phys.\ Rev.\ Lett.\  {\bf 108}, 051101 (2012)
[arXiv:1106.2000 [hep-th]].

\bibitem{fR1} 
W.~Hu and I.~Sawicki,
Phys.\ Rev.\ D {\bf 76}, 064004 (2007)
[arXiv:0705.1158 [astro-ph]].

\bibitem{fR2}
A.~A.~Starobinsky,
JETP Lett.\  {\bf 86}, 157 (2007)
[arXiv:0706.2041 [astro-ph]];

\bibitem{fR3} 
S.~A.~Appleby and R.~A.~Battye,
Phys.\ Lett.\ B {\bf 654}, 7 (2007)
[arXiv:0705.3199 [astro-ph]].

\bibitem{fR4}
S.~Tsujikawa,
Phys.\ Rev.\ D {\bf 77}, 023507 (2008)
[arXiv:0709.1391 [astro-ph]].

\bibitem{BD1} 
S.~Tsujikawa, K.~Uddin, S.~Mizuno, R.~Tavakol 
and J.~Yokoyama,
Phys.\ Rev.\ D {\bf 77}, 103009 (2008)
[arXiv:0803.1106 [astro-ph]].

\bibitem{BD2} 
R.~Gannouji, B.~Moraes, D.~F.~Mota, D.~Polarski, 
S.~Tsujikawa and H.~A.~Winther,
Phys.\ Rev.\ D {\bf 82}, 124006 (2010)
[arXiv:1010.3769 [astro-ph.CO]].

\bibitem{Nicolis} 
A.~Nicolis, R.~Rattazzi and E.~Trincherini,
Phys.\ Rev.\ D {\bf 79}, 064036 (2009)
[arXiv:0811.2197 [hep-th]].

\bibitem{Galileon1} 
C.~Deffayet, G.~Esposito-Farese and A.~Vikman,
Phys.\ Rev.\ D {\bf 79}, 084003 (2009);
[arXiv:0901.1314 [hep-th]].

\bibitem{Galileon2} 
C.~Deffayet, S.~Deser and G.~Esposito-Farese,
Phys.\ Rev.\ D {\bf 80}, 064015 (2009)
[arXiv:0906.1967 [gr-qc]].

\bibitem{GS10} 
R.~Gannouji and M.~Sami,
Phys.\ Rev.\ D {\bf 82}, 024011 (2010)
[arXiv:1004.2808 [gr-qc]].

\bibitem{DTGa1} 
A.~De Felice and S.~Tsujikawa,
Phys.\ Rev.\ Lett.\  {\bf 105}, 111301 (2010);
[arXiv:1007.2700 [astro-ph.CO]].

\bibitem{DTGa2} 
A.~De Felice and S.~Tsujikawa,
Phys.\ Rev.\ D {\bf 84}, 124029 (2011).
[arXiv:1008.4236 [hep-th]].

\bibitem{Nesseris} 
S.~Nesseris, A.~De Felice and S.~Tsujikawa,
Phys.\ Rev.\ D {\bf 82}, 124054 (2010).
[arXiv:1010.0407 [astro-ph.CO]].

\bibitem{AppleLin} 
S.~A.~Appleby and E.~V.~Linder,
JCAP {\bf 1208}, 026 (2012)
[arXiv:1204.4314 [astro-ph.CO]].

\bibitem{Totani} 
H.~Okada, T.~Totani and S.~Tsujikawa,
Phys.\ Rev.\ D {\bf 87}, 103002 (2013)
[arXiv:1208.4681 [astro-ph.CO]].

\bibitem{Barreira1}
A.~Barreira, B.~Li, A.~Sanchez, C.~M.~Baugh and S.~Pascoli,
Phys.\ Rev.\ D {\bf 87}, 103511 (2013)
[arXiv:1302.6241 [astro-ph.CO]].

\bibitem{Neveu}
J.~Neveu {\it et al.},
Astron.\ Astrophys.\  {\bf 555}, A53 (2013) 
[arXiv:1302.2786 [gr-qc]].

\bibitem{Barreira2}
A.~Barreira, B.~Li, C.~Baugh and S.~Pascoli,
JCAP {\bf 1408}, 059 (2014)
[arXiv:1406.0485 [astro-ph.CO]].

\bibitem{Renk} 
J.~Renk, M.~Zumalacarregui, F.~Montanari and A.~Barreira,
JCAP {\bf 1710}, 020 (2017)
[arXiv:1707.02263 [astro-ph.CO]].

\bibitem{Peirone} 
S.~Peirone, N.~Frusciante, B.~Hu, M.~Raveri and A.~Silvestri,
arXiv:1711.04760 [astro-ph.CO].

\bibitem{FKT10} 
A.~De Felice, R.~Kase and S.~Tsujikawa,
Phys.\ Rev.\ D {\bf 83}, 043515 (2011)
[arXiv:1011.6132 [astro-ph.CO]].

\bibitem{Zuma} 
M.~Zumalacarregui and J.~Garcia-Bellido,
Phys.\ Rev.\ D {\bf 89}, 064046 (2014).
[arXiv:1308.4685 [gr-qc]].

\bibitem{GLPV} 
J.~Gleyzes, D.~Langlois, F.~Piazza and F.~Vernizzi,
Phys.\ Rev.\ Lett.\  {\bf 114}, 211101 (2015)
[arXiv:1404.6495 [hep-th]].

\bibitem{Langlois1} 
D.~Langlois and K.~Noui,
JCAP {\bf 1602}, 034 (2016)
[arXiv:1510.06930 [gr-qc]].

\bibitem{Langlois2} 
D.~Langlois and K.~Noui,
JCAP {\bf 1607}, 016 (2016)
[arXiv:1512.06820 [gr-qc]].

\bibitem{Crisostomi16} 
M.~Crisostomi, K.~Koyama and G.~Tasinato,
JCAP {\bf 1604}, 044 (2016)
[arXiv:1602.03119 [hep-th]].

\bibitem{Gergely} 
L.~A.~Gergely and S.~Tsujikawa,
Phys.\ Rev.\ D {\bf 89}, 064059 (2014)
[arXiv:1402.0553 [hep-th]].

\bibitem{KT14} 
R.~Kase and S.~Tsujikawa,
Phys.\ Rev.\ D {\bf 90}, 044073 (2014) 
[arXiv:1407.0794 [hep-th]].

\bibitem{Koba15} 
T.~Kobayashi, Y.~Watanabe and D.~Yamauchi,
Phys.\ Rev.\ D {\bf 91}, 064013 (2015).
[arXiv:1411.4130 [gr-qc]]. 

\bibitem{Sakstein}
K.~Koyama and J.~Sakstein,
Phys.\ Rev.\ D {\bf 91}, 124066 (2015)
[arXiv:1502.06872 [astro-ph.CO]].
    
\bibitem{Mizuno}
R.~Saito, D.~Yamauchi, S.~Mizuno, J.~Gleyzes and D.~Langlois,
JCAP {\bf 1506}, 008 (2015)
[arXiv:1503.01448 [gr-qc]].

\bibitem{Babichev16} 
E.~Babichev, K.~Koyama, D.~Langlois, R.~Saito and J.~Sakstein,
Class.\ Quant.\ Grav.\  {\bf 33}, no. 23, 235014 (2016)
[arXiv:1606.06627 [gr-qc]].

\bibitem{DKT15} 
A.~De Felice, R.~Kase and S.~Tsujikawa,
Phys.\ Rev.\ D {\bf 92}, 124060 (2015)
[arXiv:1508.06364 [gr-qc]].

\bibitem{DKT15b} 
R.~Kase, S.~Tsujikawa and A.~De Felice,
JCAP {\bf 1603}, 003 (2016)
[arXiv:1512.06497 [gr-qc]].

\bibitem{KTF15} 
R.~Kase, S.~Tsujikawa and A.~De Felice,
Phys.\ Rev.\ D {\bf 93}, 024007 (2016)
[arXiv:1510.06853 [gr-qc]].

\bibitem{GW170817} 
B.~P.~Abbott {\it et al.} 
[LIGO Scientific and Virgo Collaborations],
Phys.\ Rev.\ Lett.\  {\bf 119}, 161101 (2017)
[arXiv:1710.05832 [gr-qc]].

\bibitem{Goldstein} 
A.~Goldstein {\it et al.},
Astrophys.\ J.\  {\bf 848}, no. 2, L14 (2017)
[arXiv:1710.05446 [astro-ph.HE]].

\bibitem{Abbott} 
B.~P.~Abbott {\it et al.} 
[LIGO Scientific and Virgo and Fermi-GBM and INTEGRAL Collaborations],
Astrophys.\ J.\  {\bf 848}, no. 2, L13 (2017)
[arXiv:1710.05834 [astro-ph.HE]].

\bibitem{DT11} 
A.~De Felice and S.~Tsujikawa,
JCAP {\bf 1202}, 007 (2012)
[arXiv:1110.3878 [gr-qc]].

\bibitem{GW1} 
P.~Creminelli and F.~Vernizzi,
Phys.\ Rev.\ Lett.\  {\bf 119}, 251302 (2017)
[arXiv:1710.05877 [astro-ph.CO]].

\bibitem{GW2} 
J.~M.~Ezquiaga and M.~Zumalacarregui,
Phys.\ Rev.\ Lett.\  {\bf 119}, 251304 (2017)
[arXiv:1710.05901 [astro-ph.CO]].

\bibitem{GW3} 
J.~Sakstein and B.~Jain,
Phys.\ Rev.\ Lett.\  {\bf 119}, 251303 (2017)
[arXiv:1710.05893 [astro-ph.CO]].

\bibitem{GW4} 
T.~Baker, E.~Bellini, P.~G.~Ferreira, M.~Lagos, J.~Noller and I.~Sawicki,
Phys.\ Rev.\ Lett.\  {\bf 119}, 251301 (2017)
[arXiv:1710.06394 [astro-ph.CO]].

\bibitem{GW5} 
L.~Amendola, M.~Kunz, I.~D.~Saltas and I.~Sawicki,
Phys.\ Rev.\ Lett.\  {\bf 120}, 131101 (2018)
[arXiv:1711.04825 [astro-ph.CO]].
  
\bibitem{GW6} 
M.~Crisostomi and K.~Koyama,
Phys.\ Rev.\ D {\bf 97}, 084004 (2018)
[arXiv:1712.06556 [astro-ph.CO]].

\bibitem{GW7} 
Y.~Gong, E.~Papantonopoulos and Z.~Yi,
arXiv:1711.04102 [gr-qc].

\bibitem{Lom1} 
L.~Lombriser and A.~Taylor,
JCAP {\bf 1603}, 031 (2016)
[arXiv:1509.08458 [astro-ph.CO]].

\bibitem{Lom2} 
L.~Lombriser and N.~A.~Lima,
Phys.\ Lett.\ B {\bf 765}, 382 (2017)
[arXiv:1602.07670 [astro-ph.CO]].

\bibitem{Babi17} 
E.~Babichev, C.~Charmousis, G.~Esposito-Farese and A.~Lehebel,
arXiv:1712.04398 [gr-qc].

\bibitem{building} 
J.~Gleyzes, D.~Langlois, F.~Piazza and F.~Vernizzi,
JCAP {\bf 1308}, 025 (2013)
[arXiv:1304.4840 [hep-th]].

\bibitem{KaseIJ} 
R.~Kase and S.~Tsujikawa,
Int.\ J.\ Mod.\ Phys.\ D {\bf 23}, 1443008 (2014)
[arXiv:1409.1984 [hep-th]].

\bibitem{Bellini} 
E.~Bellini and I.~Sawicki,
JCAP {\bf 1407}, 050 (2014)
[arXiv:1404.3713 [astro-ph.CO]].

\bibitem{Arkani} 
N.~Arkani-Hamed, H.~C.~Cheng, M.~A.~Luty and S.~Mukohyama,
JHEP {\bf 0405}, 074 (2004)
[hep-th/0312099].

\bibitem{EFTCAMB1} 
B.~Hu, M.~Raveri, N.~Frusciante and A.~Silvestri,
Phys.\ Rev.\ D {\bf 89}, 103530 (2014)
[arXiv:1312.5742 [astro-ph.CO]].

\bibitem{EFTCAMB2} 
M.~Raveri, B.~Hu, N.~Frusciante and A.~Silvestri,
Phys.\ Rev.\ D {\bf 90}, 043513 (2014)
[arXiv:1405.1022 [astro-ph.CO]].

\bibitem{Pogosian} 
L.~Pogosian and A.~Silvestri,
Phys.\ Rev.\ D {\bf 94}, 104014 (2016)
[arXiv:1606.05339 [astro-ph.CO]].

\bibitem{Jimenez}
J.~B.~Jimenez, F.~Piazza and H.~Velten,
Phys.\ Rev.\ Lett.\  {\bf 116}, 061101 (2016)
[arXiv:1507.05047 [gr-qc]].

\bibitem{Sakstein1} 
J.~Sakstein,
Phys.\ Rev.\ Lett.\  {\bf 115}, 201101 (2015)
[arXiv:1510.05964 v3 [astro-ph.CO]].

\bibitem{Sakstein2} 
J.~Sakstein,
Phys.\ Rev.\ D {\bf 92}, 124045 (2015)
[arXiv:1511.01685 [astro-ph.CO]].

\bibitem{Jain15} 
R.~K.~Jain, C.~Kouvaris and N.~G.~Nielsen,
Phys.\ Rev.\ Lett.\  {\bf 116}, 151103 (2016)
[arXiv:1512.05946 [astro-ph.CO]].

\bibitem{Sakstein3} 
J.~Sakstein, H.~Wilcox, D.~Bacon, K.~Koyama and R.~C.~Nichol,
 JCAP {\bf 1607},  019 (2016)
[arXiv:1603.06368 [astro-ph.CO]].

\bibitem{Cris2} 
M.~Crisostomi and K.~Koyama,
Phys.\ Rev.\ D {\bf 97}, no. 2, 021301 (2018)
[arXiv:1711.06661 [astro-ph.CO]].

\bibitem{Langlois17} 
D.~Langlois, R.~Saito, D.~Yamauchi and K.~Noui,
Phys.\ Rev.\ D {\bf 97}, 061501 (2018)
[arXiv:1711.07403 [gr-qc]].
  
\bibitem{Dima17} 
A.~Dima and F.~Vernizzi,
arXiv:1712.04731 [gr-qc].

\bibitem{BGP} 
L.~Heisenberg, R.~Kase and S.~Tsujikawa,
Phys.\ Lett.\ B {\bf 760}, 617 (2016)
[arXiv:1605.05565 [hep-th]].

\bibitem{ADM} 
R.~L.~Arnowitt, S.~Deser and C.~W.~Misner,
Phys.\ Rev.\  {\bf 116}, 1322 (1959).

\bibitem{Koyama} 
A.~De Felice, K.~Koyama and S.~Tsujikawa,
JCAP {\bf 1505}, 058 (2015)
[arXiv:1503.06539 [gr-qc]].

\bibitem{Scherrer} 
R.~J.~Scherrer,
Phys.\ Rev.\ Lett.\  {\bf 93}, 011301 (2004).

\bibitem{kinetic1} 
D.~Giannakis and W.~Hu,
Phys.\ Rev.\ D {\bf 72}, 063502 (2005)
[astro-ph/0501423].

\bibitem{kinetic2} 
F.~Arroja and M.~Sasaki,
Phys.\ Rev.\ D {\bf 81}, 107301 (2010) 
[arXiv:1002.1376 [astro-ph.CO]].

\bibitem{quasi1} 
B.~Boisseau, G.~Esposito-Farese, D.~Polarski and A.~A.~Starobinsky,
Phys.\ Rev.\ Lett.\  {\bf 85}, 2236 (2000)
[gr-qc/0001066].

\bibitem{quasi2} 
A.~De Felice, T.~Kobayashi and S.~Tsujikawa,
Phys.\ Lett.\ B {\bf 706}, 123 (2011)
[arXiv:1108.4242 [gr-qc]].

\bibitem{GLV} 
J.~Gleyzes, D.~Langlois and F.~Vernizzi,
Int.\ J.\ Mod.\ Phys.\ D {\bf 23}, no. 13, 1443010 (2014)
[arXiv:1411.3712 [hep-th]].

\bibitem{Tsuji15} 
S.~Tsujikawa,
Phys.\ Rev.\ D {\bf 92}, 044029 (2015)
[arXiv:1505.02459 [astro-ph.CO]].

\bibitem{Planck15} 
P.~A.~R.~Ade {\it et al.} [Planck Collaboration],
Astron.\ Astrophys.\  {\bf 594}, A13 (2016)
[arXiv:1502.01589 [astro-ph.CO]].

\bibitem{Okumura} 
T.~Okumura {\it et al.},
Publ.\ Astron.\ Soc.\ Jap.\  {\bf 68}, no. 3, 38, 24 (2016)
[arXiv:1511.08083 [astro-ph.CO]].

\bibitem{screeningProca} 
A.~De Felice, L.~Heisenberg, R.~Kase, S.~Tsujikawa, Y.~l.~Zhang and G.~B.~Zhao,
Phys.\ Rev.\ D {\bf 93}, 104016 (2016)
[arXiv:1602.00371 [gr-qc]].

\bibitem{HKT16} 
L.~Heisenberg, R.~Kase and S.~Tsujikawa,
Phys.\ Rev.\ D {\bf 94}, 123513 (2016)
[arXiv:1608.08390 [gr-qc]].

\bibitem{Hulse} 
R.~A.~Hulse and J.~H.~Taylor,
Astrophys. \ J.\ {\bf 195}, L51-L53 (1975).

\end{thebibliography}
\end{document}